\documentclass[iop,apj]{emulateapj}
\shorttitle{Precipitation of ENAs and Induced Escape Fluxes from Mars}
\shortauthors{Lewkow and Kharchenko}

\usepackage{graphicx}
\usepackage{color}
\usepackage{ulem}
\usepackage{amsmath}
\usepackage{multirow}
\usepackage{natbib}

\begin{document}

\title{Precipitation of Energetic Neutral Atoms and Induced Non-thermal Escape Fluxes from the Martian Atmosphere}
\author{N.R. Lewkow\altaffilmark{1,2} and V. Kharchenko\altaffilmark{1,2}}

\altaffiltext{1}{Department of Physics, University of Connecticut, Storrs CT 06269}
\altaffiltext{2}{Institute for Theoretical Atomic, Molecular, and Optical Physics, Harvard-Smithsonian Center for Astrophysics, Cambridge MA 02138}


\begin{abstract}
The precipitation of energetic neutral atoms, produced through charge exchange collisions between solar wind ions and thermal atmospheric gases, is investigated for the Martian atmosphere.
Connections between parameters of precipitating fast ions and resulting escape fluxes, altitude-dependent energy distributions of fast atoms and their coefficients of reflection from the Mars atmosphere, are established using accurate cross sections in Monte Carlo simulations. 
Distributions of secondary hot atoms and molecules, induced by precipitating particles, have been obtained and applied for computations of the non-thermal escape fluxes.
A new collisional database on accurate energy-angular dependent cross sections, required for description of the energy-momentum transfer in collisions of precipitating particles and production of non-thermal atmospheric atoms and molecules, is reported with analytic fitting equations.
3D Monte Carlo simulations with accurate energy-angular dependent cross sections have been carried out to track large ensembles of energetic atoms in a time-dependent manner as they propagate into the Martian atmosphere and transfer their energy to the ambient atoms and molecules.
Results of the Monte Carlo simulations on the energy-deposition altitude profiles, reflection coefficients, and time-dependent atmospheric heating, obtained for the isotropic hard sphere and anisotropic quantum cross sections, are compared.
Atmospheric heating rates, thermalization depths, altitude profiles of production rates, energy distributions of secondary hot atoms and molecules, and induced escape fluxes have been determined.
\end{abstract}

\keywords{atmospheric effects - atomic processes - astronomical databases: miscellaneous - planets and satellites: atmospheres – scattering}

\section{Introduction}

The evolution of planetary atmospheres is governed, in the simplest of terms, by energy input, transfer, and output.
In planetary bodies without intrinsic magnetic fields, large amounts of energy may be supplied by {\color{black} solar wind} ions into the atmosphere. 
Precipitating ions can capture electrons in collisions with atmospheric gas and very quickly become {\color{black}energetic neutral atoms (}ENAs{\color{black})} which penetrate deeply into the atmosphere before transferring their energy to the thermal gases present. 
It was estimated that ENA precipitation delivers 10$^9$ eV cm$^{-2}$ s$^{-1}$ to the atmosphere of Mars and is comparable to the energy input from EUV photons at solar minimum \citep[]{Kallio:1997}.  
Loss of neutral planetary atmospheres occurs through both thermal (Jeans) escape and non-thermal energy transfer processes, leading to atomic and molecular escape. 
Significant numbers of atmospheric non-thermal processes are induced by precipitating {\color{black} solar wind} ions. 
While thermal escape on Mars is efficient only for atomic and molecular hydrogen, the non-thermal energy transfer and escape may be the dominant source for evolution of heavier atmospheric constituents \citep[]{Hunten:1982,Johnson:2008}. 

The atmosphere of Mars has been the focus of investigations of planetary atmospheres for a long time, in particular analysis of its current and past compositions which sheds light on the loss of liquid water which is thought to have once existed on the surface of the planet \citep[]{Owen:1977,Krasnopolsky:2002,Shematovich:2007,Lammer:2013}.  
Previously calculated thermal and non-thermal escape rates of hydrogen, as well as sputtering and ion pickup, have led to estimates of an entire ocean of water with global mean depth of 30 m being lost on Mars in the past 3.8 billion years \citep[]{Krasnopolsky:2002}. 
Kinetics and energy relaxation involved in collisions between fast and thermal atoms are fundamentally important for the escape process and thus also on atmospheric evolution \citep[]{Kharchenko:1997,Bovino:2011,Fox:2014}. 
Previous works have looked at effects of {\color{black}solar wind} protons precipitating into the atmosphere of Mars using both isotropic {\color{black} hard sphere} and angular dependent forward peaked cross sections \citep[]{Kallio:2001}, as well as with accurate quantum mechanical cross sections \citep[]{Shematovich:2004,Krestyanikova:2005,Johnson:2008,Fox:2014}, but accurate energy-angular dependent cross sections have never been fully used to study non-thermal, atom-atom and atom-molecule, energy transfer and induced escape fluxes in a planetary atmosphere.  
Precipitating ENAs are created through {\color{black}charge exchange} collisions between {\color{black}solar wind} ions and atmospheric gases in the Martian atmosphere and in this work we consider these ENAs as a source for non-thermal atomic and molecular escape and compare the ENA induced escape to previously reported escape fluxes. 

The precipitation of ENAs into planetary atmospheres can be an efficient source of atmospheric heating as well as a production mechanism for {\color{black}secondary hot} atoms and molecules. 
{\color{black}Secondary hot} atoms and molecules created by ENAs essentially have non-thermal distributions and contribute significantly to total planetary escape fluxes. 
Nascent ENAs created through {\color{black}charge exchange} collisions between {\color{black}solar wind} ions and atmospheric gases maintain the vast majority of the {\color{black}solar wind} ions velocity and thus have significantly large energies, ranging from hundreds of eV/amu to several keV/amu \citep[]{Reeves:2013}. 
As the nascent ENAs precipitate through the planetary atmosphere, their energy is transferred, via elastic and inelastic collisions, to the atmospheric gases with major constituents being H, He, O, Ar, H$_{2}$, N$_{2}$, CO, and CO$_{2}$ \citep[]{Krasnopolsky:2002}. 
Extremely forward peaked differential cross sections \citep[]{Lewkow:2012} for keV collisions result in relatively small energy transfer per collision. 
This leads to several thousand collisions and deep penetration into the planetary atmosphere before thermalizing. 
Modeling of energy deposition altitude profiles requires realistic descriptions of energy transfer and thus accurate differential and total cross sections for binary collisions. 

Anisotropic {\color{black}quantum mechanical} differential cross sections, unlike isotropic {\color{black}hard sphere} approximations, are extremely forward peaked for center of mass collision energies above 1 eV. 
We have calculated with high accuracy a majority of atom-atom collision cross sections.
At the same time, ab initio calculations of atom-molecule cross sections at keV energies, such as atomic collisions with CO$_{2}$ molecules, are not realistic and semi-empirical methods should be applied.
Unknown cross sections of atom-molecule collisions between ENAs and some species of the Mars atmosphere were treated using an angular-energy dependent scaling method to provide reasonable forward peaked differential cross sections as well as integrated total cross sections.
These scaling cross sections are useful in the atmosphere of Mars where CO, CO$_{2}$, H$_{2}$, and N$_{2}$ are large constituents and accurate {\color{black}quantum mechanical}, ab initio computations at keV/amu collision energies look as very formidable problems. 
All collisions between ENAs and these atmospheric molecules utilize the scaling cross sections, while all known atom-atom collisions (H+H, He+H, He+He, He+O) use computed ab initio quantum mechanical cross sections in this work. 
Computed cross sections as well as results of quantum scaling have been verified with available experimental data \citep[]{Gao:1989,Newman:1986,Nitz:1987,Smith:1996,Schafer:1987}.

Through use of {\color{black}quantum mechanical} and scaling cross sections, accurate time-dependent calculation of ENA transport, momentum transfer energy loss, {\color{black}secondary hot} atomic and molecular production and escape was carried out using three-dimensional {\color{black}Monte Carlo} simulations with large ensembles of test particles. 
Direct connections between the mechanisms of energy deposition and the intensities of induced escape fluxes for neutral atoms and molecules has been established using realistic cross sections, simulating {\color{black}quantum mechanical} binary collisions, combined with classical {\color{black}Monte Carlo} transport. 
Energy distributions for both thermalizing and escaping ENAs were found for ensembles of mono-energetic precipitating ENAs as well as realistic ENA energy distributions which reflect the actual energy distributions in {\color{black}solar wind} ions \citep[]{Reeves:2013}. 
Energy-deposition and escape flux comparisons between realistic anisotropic cross sections and isotropic {\color{black}hard sphere} models were made to further analyze differences in thermalization parameters between the two cross section models. 

Details on determination of both differential and total cross sections used in this study are given in section 2. 
Section 3 discusses production rates of ENAs in the upper atmosphere of Mars for the different atmosphere compositions appropriated to low, high, and mean solar activity while section 4 examines all details of the {\color{black}Monte Carlo} simulation developed for this work. 
Results obtained by the simulations and the implications for atmospheric evolution are covered in sections 5. 
Concluding remarks follow. 

\section{Cross Sections}

Accurate anisotropic cross sections are crucial for realistic simulation of energy deposition, momentum-transfer, and induced {\color{black}secondary hot (}SH{\color{black})} atomic and molecular production rates \citep[]{Kharchenko:1997,Shematovich:2004,Krestyanikova:2005,Johnson:2008,Bovino:2011,Lewkow:2012,Fox:2014}.
Quantum mechanical {\color{black}(QM)} elastic cross sections for H+H, H+He, He+He, and He+O were previously calculated from 0.01 eV to 10 keV center of mass collisional energies using ab initio interaction potentials and standard partial wave methods \citep[]{Lewkow:2012}. 
Results of calculations are in excellent agreement with available data from laboratory experiments \citep[]{Gao:1989,Newman:1986,Nitz:1987,Smith:1996,Schafer:1987}. 
Only elastic channels were considered for calculation of these atom-atom collisions as elastic cross sections dominant inelastic cross sections within this energy range \citep[]{Fridman:2012}. 
All other atom-atom and atom-molecule collisions, for which accurate interaction electronic energy surfaces are not available, utilized the newly developed scaling differential cross sections, described below, to obtain realistic energy-angular dependent cross sections.
  
\begin{figure}
	\vspace{20pt}
  \plotone{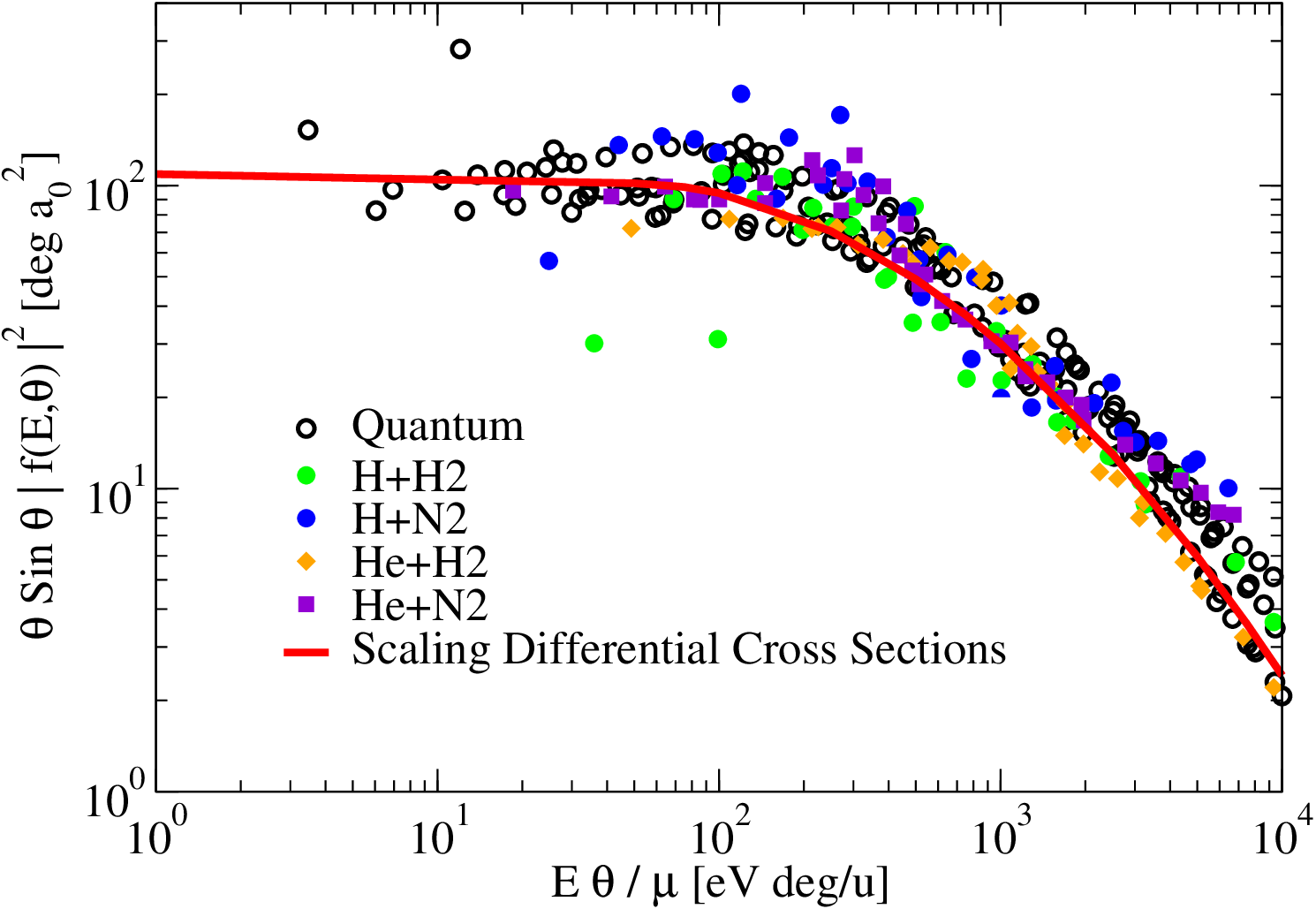}
  \caption{{\footnotesize Scaling differential cross section shown along with computed quantum mechanical H+H, He+H, He+He, and He+O differential cross sections \citep[]{Lewkow:2012}, given by black circles, as well as experimental H+H$_{2}$, H+N$_{2}$, He+H$_{2}$, and He+N$_{2}$ differential cross sections \citep[]{Newman:1985,Newman:1986}. The scaling differential cross section fit is shown with a solid red line. All data are displayed in reduced velocity coordinates \cite{Lewkow:2012}.}}
  \label{UniversalExpt}
\end{figure}

The newly developed procedure for scaling differential cross sections has been used to determine unknown atom-atom and atom-molecule collisions \citep[]{Lewkow:2012}. 
Unknown cross sections for complicated atom-molecule collisions, such as O+CO$_{2}$, have been obtained using the reduced coordinate scaling procedure, modified to be dependent on velocity instead of energy, effectively incorporating the reduced mass of the system into the scaling \citep[]{Lewkow:2012}.
The scaling procedure requires well known ``basic'' atom-atom and atom-molecule differential cross section data for several collision species over a relatively broad interval of collision energies. 
This ``basic'' set of cross sections have to be plotted in scaling variables $y$ vs $x$:
\begin{equation}
y = \frac{\theta \, \sin \theta}{\gamma} \, |f(E,\theta)|^2, \qquad x = \frac{E \theta}{\mu},
\end{equation}
where $\theta$ and $E$ are the scattering angle and collision energy in the {\color{black}center of mass} frame, $\mu$ is the reduced mass of the system, {\color{black}$\gamma$} is a scaling constant to differentiate atom-atom and atom-molecule collisions, and $|f(E,\theta)|^2$ is the center of mass differential cross section.
When displayed in this fashion, differential cross sections for several different collision species lie upon a single curve which may then be fit to a simple quadratic for large values of $x$, and a linear function for small values of $x$. 
Both fitting functions are applied to data in log-log scale.
The scaling differential cross section may then be written:
\begin{equation}
\begin{split}
|f(E,\theta)|^2 = \frac{\gamma}{\theta \, \sin \theta} \exp \left[ C_{1} \left( \log \frac{E \theta}{\mu} \right)^2 + C_{2} \log \frac{E \theta}{\mu} + C_{3} \right], \\
\frac{E \theta}{\mu} \ge x_{0}, \\
|f(E,\theta)|^2 = \frac{\gamma}{\theta \, \sin \theta} \, C_{4} \, \exp \left[ C_{5} + C_{6} \log \frac{E \theta}{\mu}  \right] + C_{7}, \\
\frac{E \theta}{\mu} < x_{0},
\label{uni_amp}
\end{split}
\end{equation}
where the fitting parameters $C_{1}$, $C_{2}$, $C_{3}$, $C_{4}$, $C_{5}$, $C_{6}$, and $C_{7}$ were found to be -0.13, 1.00, 2.70, 10.0, 2.04, -0.03, and 32.3 respectively, and the cutoff parameter is $x_{0} = 50.12$.
The scaling constant is taken as $\gamma = 1$ for atom-atom collisions and an empirical value of $\gamma = 1.4$ for atom-molecule collisions {\color{black}which was employed to scale the atom-molecule collisions to lie upon the same curve as the atom-atom collisions}.
{\color{black}The scaling differential cross section in Equation \ref{uni_amp} differs from the one presented previously in \cite{Lewkow:2012} as the current scaling differential cross section includes both a low energy/angle, linear fit as well as a high energy/angle quadratic fit, both fits being carried out in log-log coordinates.
The piecewise fitting of Equation \ref{uni_amp} allows for accurate differential cross sections over a wide range of energies and scattering angles not previously available from the scaling differential cross sections presented in \cite{Lewkow:2012}.}

\begin{figure}
	\vspace{20pt}
  \plotone{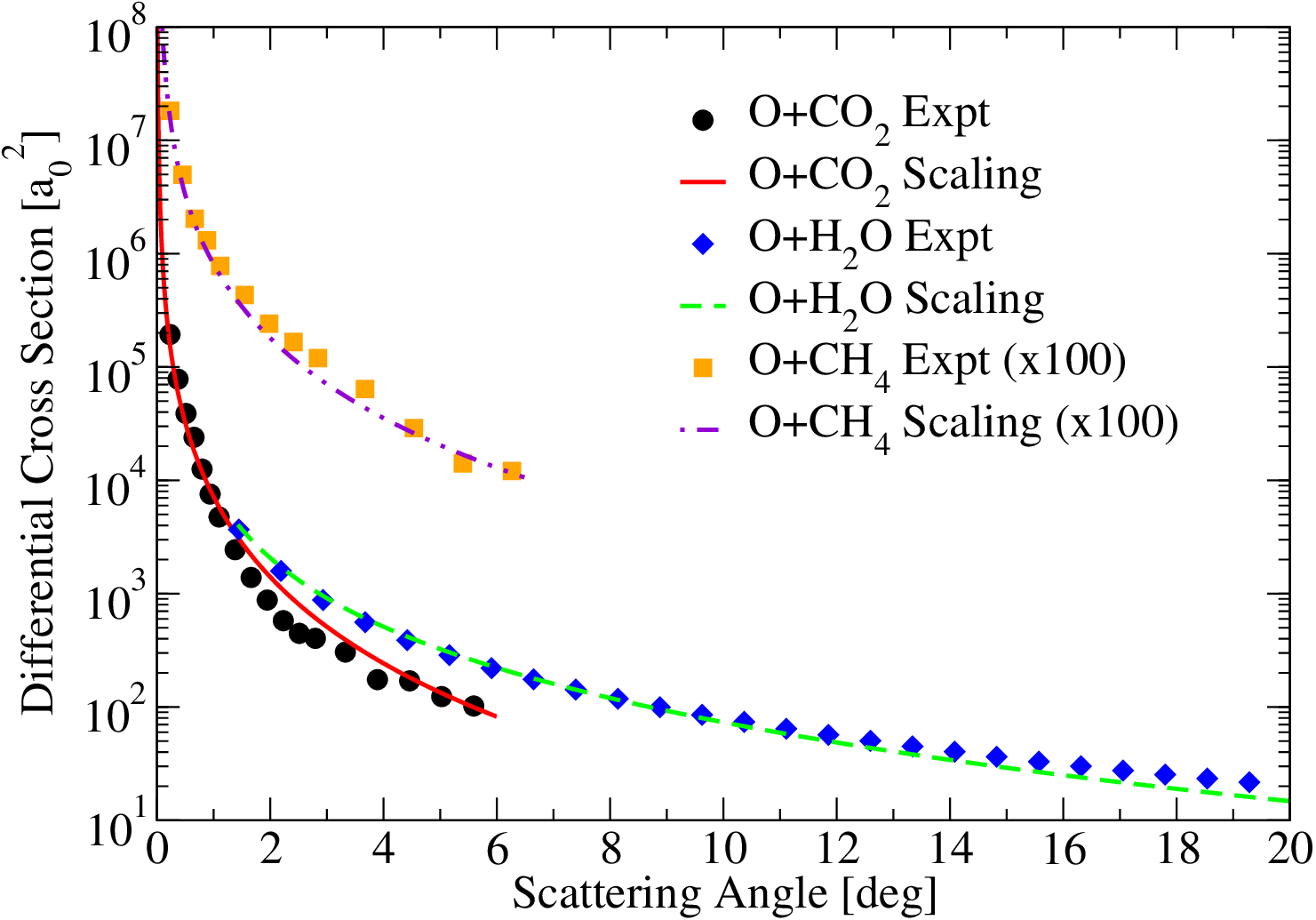}
	\caption{{\footnotesize Experimental differential cross sections for collisions of 1.5 keV O+CO$_{2}$, 100 eV O+H$_{2}$O, and 500 eV O+CH$_{4}$ shown as circles, diamonds and squares \citep[]{Smith:1996}. 
Predicted scaling differential cross sections shown as solid and dashed lines.
All collision energies, scattering angles, and differential cross sections are shown in the lab frame.}}
  \label{ScalingOCO2}
\end{figure}

{\color{black}The scaling differential cross section in Equation \ref{uni_amp} is easily utilized for complex scattering problems involving both atom-atom collisions and atom-molecules collisions through insertion of the reduced mass of the system.
For example, an important, complex collision in the atmosphere of Mars involves O+CO$_{2}$ which may be calculated using Equation \ref{uni_amp} with a reduced mass of $\mu = m_{O} m_{CO_{2}}/(m_{O} + m_{CO_{2}}) = 11.73$ amu where $m_{O}$ and $m_{CO_{2}}$ are the masses of the oxygen atom and carbon dioxide molecule respectively.
The next step in the application of Equation \ref{uni_amp} to O+CO$_{2}$ collisions is evaluation of the cutoff parameter $x_{0}$.
At a fixed collision energy $E$ we can determine specific formulas for the differential cross section as a function of the scattering angle $\theta$: if the scattering angle is inside the interval $0 \le \theta \le \mu x_{0}/E = 588.1$ eV/E degrees, the lower part of Equation \ref{uni_amp} which utilizes constants C$_{4}$--C$_{7}$ is used.
If instead $588.1 \mbox{ ev/E}=\mu x_{0}/E \leq \theta \leq 180$ degrees then the upper part of Equation \ref{uni_amp} which utilizes constants C$_{1}$--C$_{3}$ is used. 
The considered numerical example will be employed later in Figure \ref{ScalingOCO2} to plot the differential cross sections of O+CO$_{2}$ collisions as a function of the scattering angle $\theta$ and to compare theoretical results with experimental data.
}
The units needed for {\color{black}Equation} \ref{uni_amp} are E[eV], $\theta$[deg], and $\mu$[amu], where the energy and scattering angle are both taken in the {\color{black}center of mass} frame and the resulting differential cross section has units of [a$_{0}^2$].

The ``basic'' differential cross sections determined in ab initio {\color{black}QM} calculations and verified with available experimental and theoretical data for atom-atom collisions H+H, He+H, He+He, and He+O \citep[]{Lewkow:2012}, as well as experimental atom-molecule collisions H+H$_{2}$, H+N$_{2}$, He+H$_{2}$, and He+N$_{2}$ \citep[]{Newman:1985,Newman:1986}, all of which are common collisions in the atmosphere of Mars, are shown in Figure \ref{UniversalExpt} along with the scaling differential cross section fit shown in red.
Figure \ref{UniversalExpt} demonstrates how differential cross sections from a variety of different collision species form a distinctive grouping when plotted in reduced velocity coordinates. 
The scaling differential cross section fit in Figure \ref{UniversalExpt} is on the same order of magnitude as all experimental and ab initio differential cross section for all energies, and for higher energies within a factor of 5 from the experimental and ab initio data.  
Although the difference between experimental/quantum differential cross sections and the scaling differential cross section become larger at lower collision energies where quantum wave effects greatly affect scattering parameters, the scaling differential cross section allows energy-angular dependent cross sections to be obtained which are much more accurate than commonly used isotropic, hard sphere {\color{black}(HS)} cross sections. 

To demonstrate the potential power of the scaling procedure, differential cross sections for collisions between O atoms and the molecules CO$_{2}$, H$_{2}$O, and CH$_{4}$ have been calculated using this new scaling method and compared to experimental differential cross sections in Figure \ref{ScalingOCO2} for lab frame energies of 1.5 keV, 100 eV, and 500 eV respectively \citep[]{Smith:1996}.
Predicted and experimental cross sections are in an excellent agreement, taking into account that O+CO$_{2}$, O+H$_{2}$O, and O+CH$_{4}$ differential cross sections are not included in the ``basic'' cross sections.  
These specific collisions are important in different astrophysical environments, such as planetary, {\color{black}satellite}, and cometary atmospheres. 
For example, H$_{2}$O is a major component of cometary atmospheres as well as comprising $\sim$ 0.25\% of the Earth's atmosphere \citep[]{Wallace:2006}, and CH$_{4}$ is a major component in the atmosphere of Saturn's moon Titan \citep[]{Cui:2012}. 
Figure \ref{ScalingOCO2} shows the experimental and scaling differential cross sections with excellent agreement for lab frame energies ranging from 100 eV to 1.5 keV and lab frame scattering angles up to 20$^{\circ}$.

It is often the case that atom-molecule collisions may occur {\color{black}through} inelastic channels, stimulating rotational and vibrational excitations in the molecular species. 
In this study, inelastic atom-molecule collisions all utilize the scaling differential cross section, which was constructed using elastic {\color{black}QM} cross sections as well as experimental atom-molecule cross sections, which naturally include both elastic and inelastic channels.

\begin{figure}
	\vspace{20pt}
	\includegraphics[width=.85\linewidth]{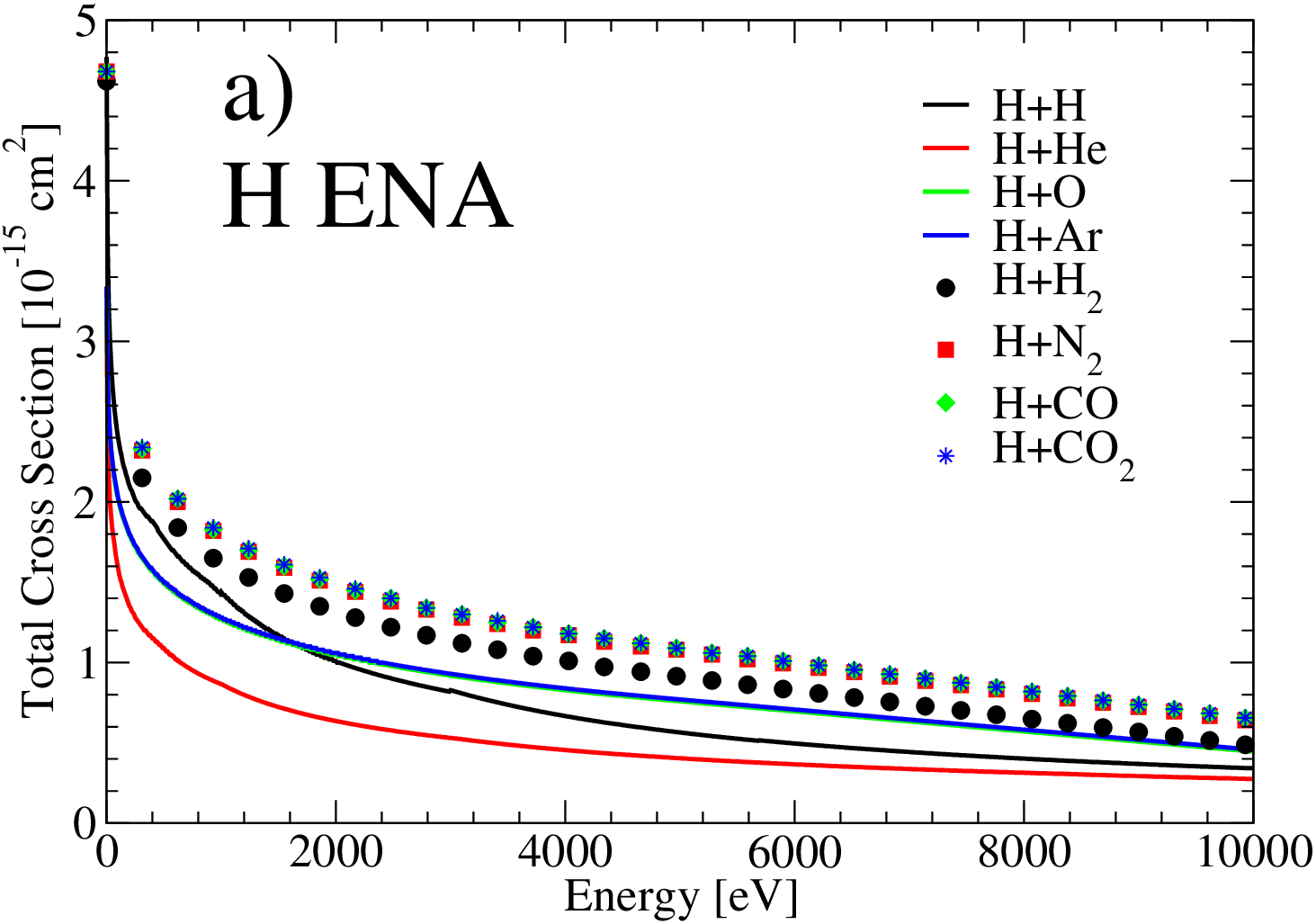}\hfill
	\vspace{30pt}
	\includegraphics[width=.85\linewidth]{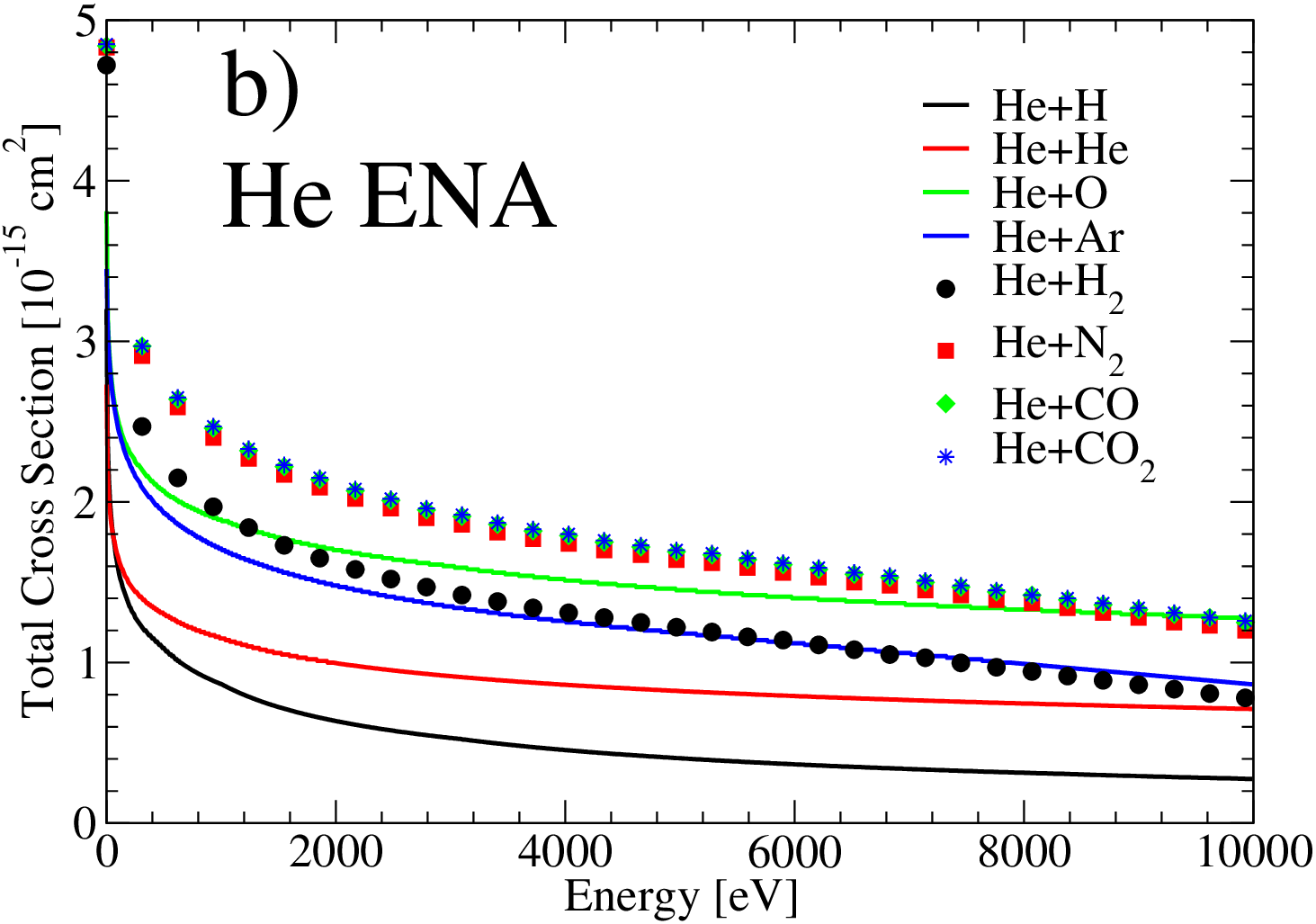} \\
  \caption{\footnotesize Total cross sections obtained from numeric integration of the scaling differential cross sections as well as {\color{black}QM} partial wave analysis. a) displays total cross sections for hydrogen ENAs with the atmospheric constituents of Mars while b) displays the same data for helium ENAs. Collision energies are shown in the {\color{black}center of mass} frame.}
  \label{All_TCS}
\end{figure}

Total cross sections for unknown atom-molecule collisions were obtained through numeric integration of the scaling differential cross sections given by {\color{black}Equation} \ref{uni_amp}. 
The numerical integration bounds used for all energies were set as $\theta_{min}$ = 0.01 deg and $\theta_{max}$ = 170 deg as {\color{black}Equation} \ref{uni_amp} {\color{black}goes} to infinity as $\theta$ goes to 0 deg and 180 deg. 
The choice of integration bounds was validated by comparing integrated scaling total cross sections with accurate computed {\color{black}QM} cross sections for collisions of H+H, He+H, He+He, and He+O. 
An average error of scaled cross section predictions is about 23\% over {\color{black}center of mass} collision energies from 1 eV to 10 keV. 
Figure \ref{All_TCS} shows the total cross sections used in this study. 
{\color{black}Additionally, all scaling total cross sections were fit to the analytic form
\begin{equation}
\sigma(E) = \sigma_{0} \left( \frac{E_{0}}{E} \right)^{\alpha}
\end{equation}
where $\sigma_{0}$ and $\alpha$ are fitting constants, $E_{0}$ is 1 keV, and $E$ is the center of mass collision energy \citep[]{Lewkow:2012}.
Table \ref{ScalingTCSFits} gives fitting constants for all collisions which utilized the scaling cross sections over several collision energy intervals.}
It should again be noted that the collisions between H+H, He+H, He+He, and He+O all employed previously calculated quantum mechanical cross sections verified with available experimental data \citep[]{Gao:1989,Newman:1986,Nitz:1987,Smith:1996,Schafer:1987} while all other collisions utilized numeric integration of the scaling differential cross sections to obtain the total cross sections. 

\begin{table*}
\begin{center}
  \caption{Scaling total cross section fitting parameters\label{ScalingTCSFits}}
  \begin{tabular}{ | c | l | c | c | c | }
\tableline
\tableline
     \textbf{Collision} & \textbf{Energy [eV]} & \textbf{$\sigma_{0}$ [10$^{-15}$ cm$^2$]} & \textbf{$\alpha$ [10$^{-2}$]} & \textbf{Max \% Error} \\
\tableline
              			& 0.1--1  			& 2.66 & 2.86 & 1.0 \\
              			& 1--10   			& 1.81 & 9.16 & 1.5 \\
\textbf{H+O}  			& 10--100 			& 1.42 & 15.2 & 1.3 \\
              			& 100--1,000    & 1.26 & 23.1 & 2.7 \\
       			  			& 1,000--10,000 & 1.41 & 39.7 & 2.4 \\
\tableline
              			& 0.1--1 				& 2.68 & 2.81 & 1.0 \\
              			& 1--10 				& 1.82 & 9.06 & 1.5 \\
\textbf{H+Ar} 			& 10--100 			& 1.43 & 15.2 & 1.3 \\
              			& 100--1,000 		& 1.27 & 23.0 & 2.7 \\
       			  			& 1,000--10,000 & 1.42 & 39.3 & 2.3 \\
\tableline
              			& 0.1--1 				& 3.52 & 3.52 & 1.2 \\
              			& 1--10 				& 2.33 & 10.2 & 1.4 \\
\textbf{H+H$_{2}$} 	& 10-100 				& 1.84 & 16.1 & 1.4 \\
              		 	& 100--1,000 		& 1.61 & 25.0 & 3.2 \\
       			  			& 1,000--10,000 & 1.84 & 45.1 & 2.8 \\
\tableline
              			& 0.1--1 				& 3.75 & 2.82 & 1.0 \\
              			& 1--10 				& 2.55 & 9.08 & 1.5 \\
\textbf{H+N$_{2}$} 	& 10-100 				& 2.00 & 15.2 & 1.3 \\
              		 	& 100--1,000 		& 1.78 & 23.0 & 2.7 \\
       			  			& 1,000-10,000 	& 1.98 & 39.4 & 2.3 \\
\tableline
              			& 0.1--1 				& 3.75 & 2.82 & 1.0 \\
              			& 1--10 				& 2.55 & 9.08 & 1.5 \\
\textbf{H+CO} 	    & 10--100 			& 2.00 & 15.2 & 1.3 \\
              		 	& 100--1,000 		& 1.78 & 23.0 & 2.7 \\
       			  			& 1,000-10,000 	& 1.98 & 39.4 & 2.3 \\
\tableline
              			& 0.1--1 				& 3.75 & 2.80 & 1.0 \\
              			& 1--10 				& 2.56 & 9.05 & 1.5 \\
\textbf{H+CO$_{2}$} & 10--100 			& 2.00 & 15.1 & 1.3 \\
              		 	& 100--1,000 		& 1.78 & 22.9 & 2.7 \\
       			  			& 1,000--10,000 & 1.99 & 39.2 & 2.3 \\
\tableline
              			& 0.1--1 				& 2.96 & 1.86 & 0.1 \\
              			& 1--10 				& 2.45 & 5.09 & 1.5 \\
\textbf{He+Ar}      & 10--100 			& 1.87 & 11.9 & 1.3 \\
              		 	& 100--1,000 		& 1.70 & 17.9 & 1.6 \\
       			  			& 1,000--10,000 & 1.79 & 26.4 & 1.1 \\
\tableline
              			& 0.1--1 				& 3.90 & 2.39 & 0.7 \\
              			& 1--10 				& 2.75 & 8.10 & 1.6 \\
\textbf{He+H$_{2}$} & 10--100 			& 2.14 & 14.4 & 1.3 \\
              		 	& 100--1,000 		& 1.92 & 21.5 & 2.3 \\
       			  			& 1,000--10,000 & 2.10 & 35.2 & 1.9 \\
\tableline
              			& 0.1--1 				& 4.14 & 1.87 & 0.1 \\
              			& 1--10 				& 3.41 & 5.17 & 1.5 \\
\textbf{He+N$_{2}$} & 10--100 			& 2.61 & 11.9 & 1.3 \\
              		 	& 100--1,000 		& 2.36 & 18.0 & 1.6 \\
       			  			& 1,000--10,000 & 2.50 & 26.6 & 1.1 \\
\tableline
              			& 0.1--1 				& 4.14 & 1.87 & 0.1 \\
              			& 1--10 				& 3.41 & 5.17 & 1.5 \\
\textbf{He+CO}      & 10--100 			& 2.61 & 11.9 & 1.3 \\
              		 	& 100--1,000 		& 2.36 & 18.0 & 1.6 \\
       			  			& 1,000--10,000 & 2.50 & 26.6 & 1.1 \\
\tableline
              			& 0.1--1 				& 4.14 & 1.86 & 0.1 \\
              			& 1--10 				& 3.44 & 5.03 & 1.5 \\
\textbf{He+CO$_{2}$}& 10--100 			& 2.63 & 11.8 & 1.3 \\
              		 	& 100--1,000 		& 2.39 & 17.8 & 1.6 \\
       			  			& 1,000--10,000 & 2.52 & 26.3 & 1.0 \\
\tableline
\end{tabular}
\end{center}
\tablecomments{Fitting parameters are shown for all collisions which utilized the scaling cross sections. Fitting parameters $\sigma_{0}$ and $\alpha$ are given for several center of mass energy intervals to ensure good fits. The maximum \% error over the given energy interval is also shown giving a measure of the fitting efficiency.}
\end{table*}

Once differential and total cross sections were obtained, a normalized scattering angle probability density, $\rho(E,\theta)$, was constructed 
\begin{equation}
\rho(E,\theta) = \frac{2 \pi}{\sigma(E)} \sin \theta \, |f(E,\theta)|^2, \qquad \int\limits_{0}^{\pi} \rho(E,\theta) \, d\theta = 1, 
\end{equation}
where $E$ is the {\color{black}center of mass} collision energy, $\theta$ is the {\color{black}center of mass} scattering angle, $|f(E,\theta)|^2$ is the {\color{black}center of mass} differential cross section, and $\sigma(E)$ is the total cross section. 
The scattering angle probability density was then used to build a cumulative scattering probability 
\begin{equation}
P(E,\theta) = \int\limits_{0}^{\theta} \rho(E,\theta') \, d\theta', \qquad P(E,\theta) \in [0,1],
\label{Cum_Prob}
\end{equation}
which gives the probability to scatter into angles less than $\theta$. 
The cumulative scattering probability $P(E,\theta)$ is a good metric for the scattering angle anisotropy of a given collision. 
For example, in the case of the H+He collision, 50\% of the cumulative probability was reached at lab frame scattering angles of 0.2 degrees, 0.1 degrees, and 0.06 degrees for lab frame collision energies of 0.5 keV, 1.5 keV, and 5 keV respectively \citep[]{Lewkow:2012}. 
\begin{figure}
	\vspace{20pt}
	\includegraphics[width=.85\linewidth]{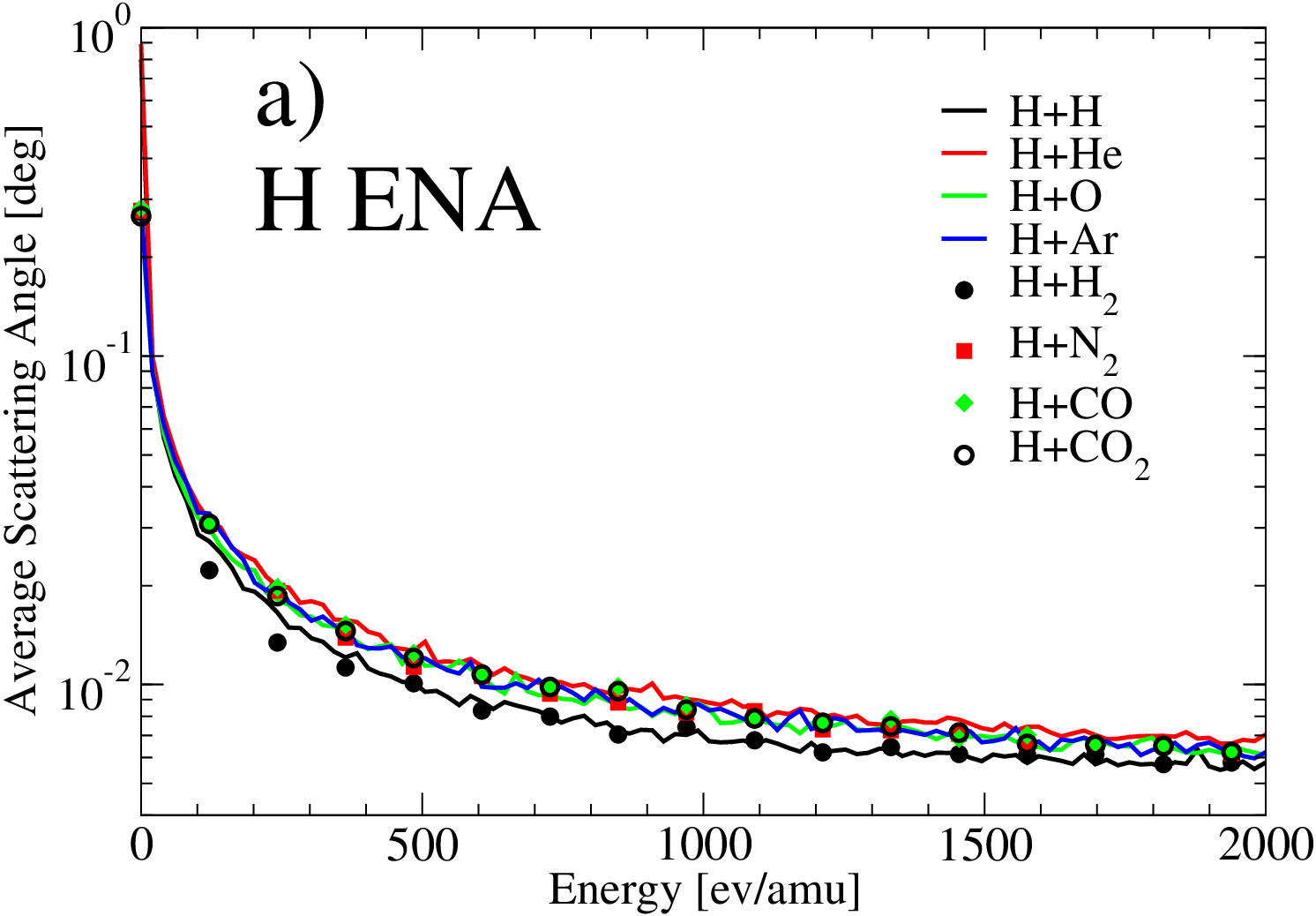}\hfill
	\vspace{30pt}
	\includegraphics[width=.85\linewidth]{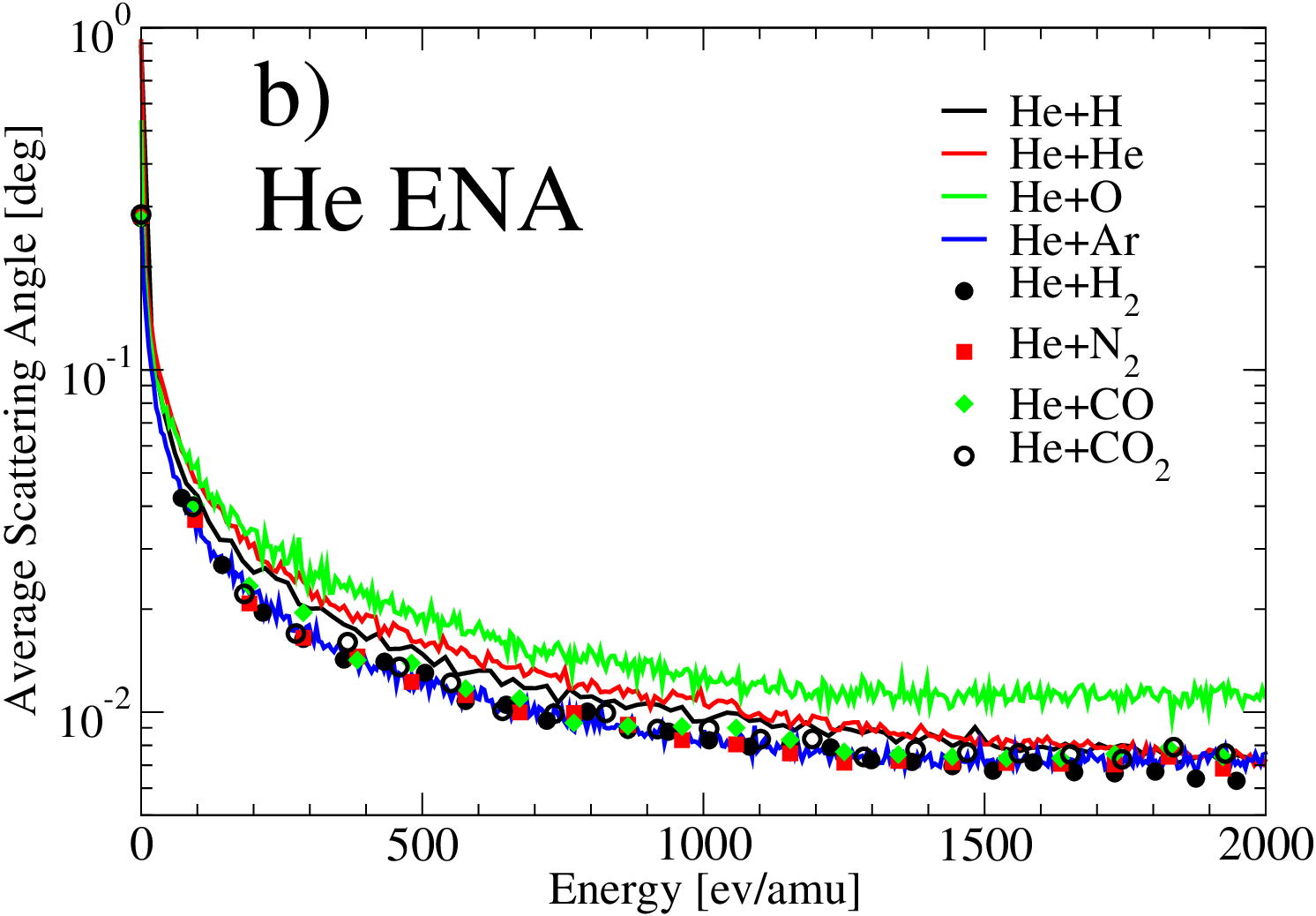} \\
  \caption{Average {\color{black}center of mass} scattering angles as a function of {\color{black}center of mass} collision energy/amu for a) H ENAs and b) He ENAs with atmospheric constituents of Mars.}
	\label{Rand_Angles}
	\vspace{10pt}
\end{figure}
In addition, numeric inversion of the cumulative probability, Equation \ref{Cum_Prob}, may also be used to generate random scattering angles for a given collision species such that $\eta=P(E,\theta_{rnd})$, where $\theta_{rnd}$ is the random scattering angle associated with a uniform random number $\eta$. 
Figure \ref{Rand_Angles} displays average scattering angles as a function of collision energy for both hydrogen and helium projectiles with Mars' atmospheric gases. 
These forward peaked, heavily anisotropic cross sections result in both deep penetration of ENAs into the planetary atmosphere as well as several thousand collisions before thermalizing with the atmospheric gases.
This effectively results in the energy deposition process involving significantly more layers of the atmospheric gas than in calculations with isotropic cross sections. 
The average scattering angles in Figure \ref{Rand_Angles} further demonstrate the forward peaked nature of these high energy collisions. 
Furthermore, the effectiveness of the scaling differential cross sections is also demonstrated in Figure \ref{Rand_Angles} as collisions which utilized this new method have average scattering angles which are similar to those obtained using {\color{black}QM} cross sections. 
It should also be noted that although these keV cross sections are extremely forward peaked, the low probable, large scattering angles are also important for transport and thermalization, as they are the main mechanism for production of relatively fast SH atoms and molecules.

In the following sections, the newly developed cross section database is utilized to model energy transfer, SH atom and molecule production and non-thermal escape for the Martian atmosphere. 

\section{The Martian Atmosphere and ENA Production}

The main mechanism for the production of ENAs in the upper Mars atmosphere is {\color{black}charge exchange (}CX{\color{black})} collisions between energetic {\color{black}solar wind (}SW{\color{black})} ions and thermal atmospheric gases \citep[]{Kallio:1997}. 
The source of the energetic ions depends on the specific astrophysical environment. 
SW ions may be a significant source of energetic ions for many planetary bodies, while planets with intrinsic magnetic fields, such as Jupiter or Earth, also have magnetospheric ions which contribute significantly to ENA production.

A statistical analysis of SW velocity data taken from 1989 through 2012 shows that the average SW energy ranges from 0.7 keV/amu to 1.5 keV/amu with the most energetic SW reaching energies of 4.2 keV/amu \citep[]{Reeves:2013}. 
Theoretical mono-energetic ensembles of SW ions as well as the realistic average energy distributions of ions in the SW plasma were used to analyze how parameters of the energy-deposition and precipitating fluxes change with initial energy distributions of SW ions and nascent ENAs. 

Different models for the neutral upper atmosphere of Mars are required for both ENA production and {\color{black}Monte Carlo (}MC{\color{black})} transport simulations with the nascent ENAs at different solar conditions. 
Three models were used in this work to determine how drastically the energy-transfer, thermalization, SH atom/molecule production, and escape fluxes are changed with different solar activity {\color{black}(SA)}. 
In this work, neutral atmosphere models representing high, low, and mean SA were employed for the atomic and molecular species H, He, O, Ar, H$_{2}$, N$_{2}$, CO, and CO$_{2}$ \citep[]{Krasnopolsky:2002}.  
While the density profiles provided by \cite{Krasnopolsky:2002} extend up to 300 km, higher altitude densities are needed for this study. 
For simplicity, exponential fits were used at these high altitudes, a method which is commonly applied in atmospheric modeling \citep[]{Sanchez:2010}.  

In addition to SW energy distributions and neutral atmosphere models for Mars, CX cross sections were also required to determine production rates of nascent ENAs. 
Accurate, energy-dependent CX cross sections were used for all required ion+atom collisions over the necessary collision energy range \citep[]{Barnett:1990,Lindsay:2005,Kusakable:2002,Greenwood:2000,Gealy:1987,Stier:1956}. 
CX collisions occurring at keV energies are extremely forward peaked \citep[]{Gao:1988,Johnson:1989,Gao:1990} so that nascent ENAs maintain nearly the same velocity magnitude and direction of its parent ion.  
The probability of charge stripping processes is very low at keV/amu collision energies and nascent ENAs, induced in {\color{black}CX} collisions, propagate through the atmosphere as high speed neutral particles. 

Although the SW ions are quickly converted to nascent ENAs, for consistency of the theoretical description, we also consider the energy loss of precipitating SW ions.
For typical SW ion velocities, the energy losses related to the ionization and excitation of atmospheric atoms and molecules are small and major energy losses occur in elastic collisions of precipitating ions with the ambient gas.
The effect of SW ion elastic collisions was investigated to determine average energy losses prior to creation of ENAs from CX collisions. 
Energy-dependent elastic cross sections for H$^+$+H \citep[]{Schultz:2008} were used for all atmospheric species in this investigation as actual elastic atom-ion cross sections for all neutral species of interest were not readily available in the literature. 
Energy loss in these ion collisions was determined using angular-energy dependent scaling differential cross sections to determine random scattering angles as described for atom-atom and atom-molecule collisions in the previous section.  
We found that the incident SW ions undergo an average of 3 elastic ion-atom collisions and lose an average of 0.1 eV from their initial energy before becoming nascent ENAs through CX collisions with atmospheric neutrals. 
These results were obtained using 100,000 particle ensembles which were initialized at 800 km using the model SW energy distribution \citep[]{Reeves:2013} and were directly incident on the planet surface with a solar zenith angle of 0 deg. 
Certainly, such small ion energy losses can be neglected taking into account typical energies of precipitating SW ions are around a few keV/amu and they're quickly converted to ENAs.
{\color{black}The results of average ion energy loss prior to ENA production may be different if accurate elastic ion-atom cross sections were used as opposed to the H$^+$+H cross sections for all species, although the final result of negligible energy loss relative to the ion energies would remain.
For example, in the extreme case where some ion-atom cross sections are 2 orders of magnitude larger than the H$^+$+H cross sections, the resulting energy loss would be an average of 10 eV before ENA production, which only amounts to 1\% and 0.3\% for hydrogen and helium ions respectively.}

\begin{figure}
	\vspace{20pt}
  \plotone{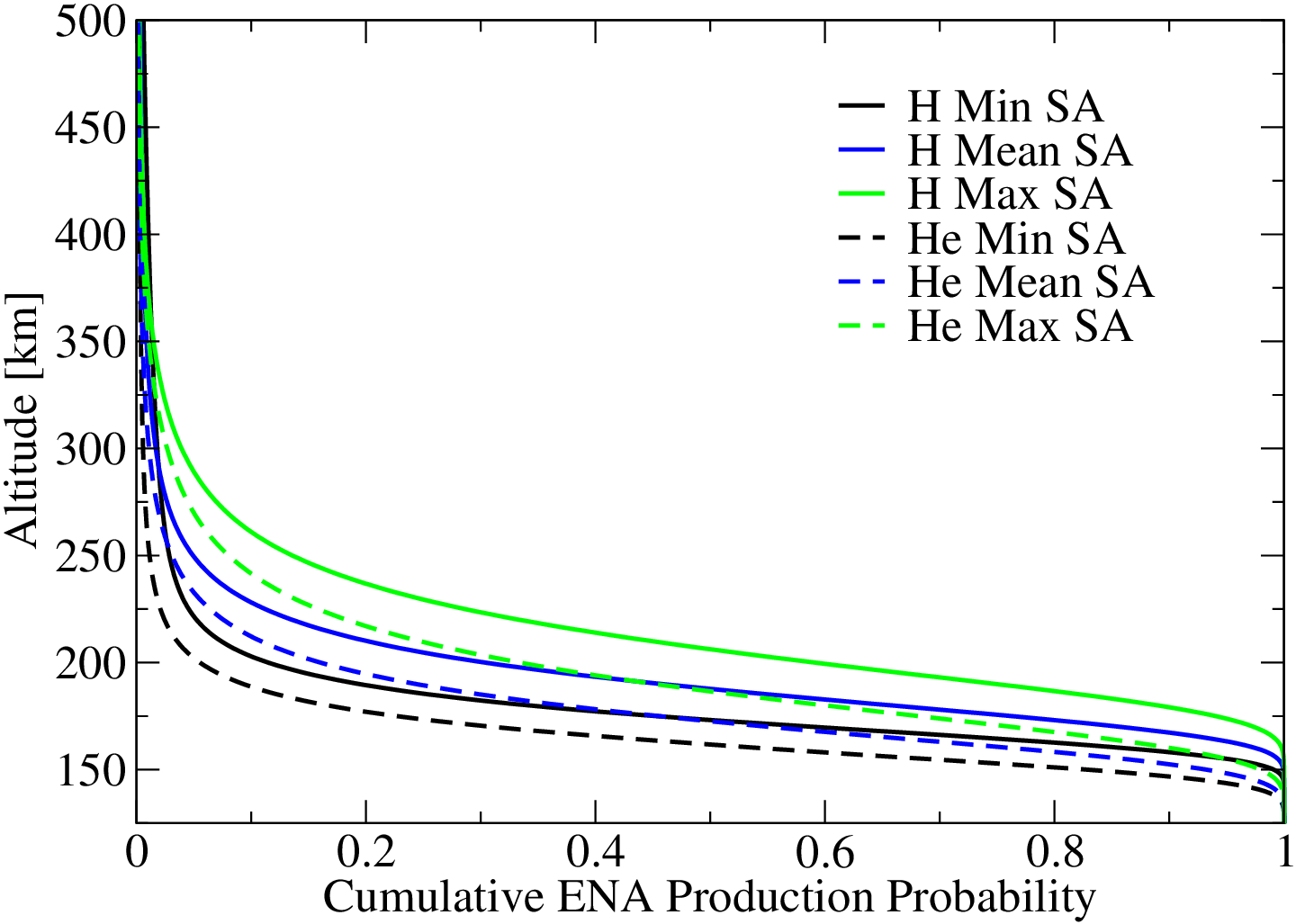}
  \caption{{\footnotesize Altitude dependent cumulative distribution function for the nascent ENA production. Both hydrogen and helium ENAs are shown with solid and dashed lines for the three different Mars atmosphere models corresponding to minimum, mean, and maximum SA.}}
  \label{ENA_Prod}
\end{figure}

Nascent ENA production fluxes, $f(z)$, were calculated using a simplified 1D atmospheric transparency integral:
\begin{equation}
\begin{split}
	f(z) = N_{sw} U_{sw} \frac{R_{0}^2}{R_{mars}^2} \exp \left( - \sum_{i} \int\limits_{z}^{\infty} \sigma_{i}^{cx} n_{i}(z') \, dz' \right) \\
\times \sum_{i} \sigma_{i}^{cx} n_{i}(z), 
\label{Trans_Int}
\end{split}
\end{equation} 
where $z$ is the altitude above Mars, $U_{sw}$ and $N_{sw}$ are the velocity and density of the SW ions at $R_{0}$ 1 AU, $R_{mars}$ is 1.5 AU, $\sigma_{i}^{cx}$ is the CX cross section for the i$^{th}$ neutral species, and $n_{i}$ is the neutral density of the i$^{th}$ neutral species at altitude $z$.
Cumulative distribution functions for ENA production altitudes, for all three atmosphere models, are shown in Figure \ref{ENA_Prod} for nascent hydrogen and helium ENAs using realistic SW velocity distributions \citep[]{Reeves:2013} and an average SW density of 5 cm$^{-3}$ at 1 AU \citep[]{Phillips:1995}. 
These cumulative distribution functions give a probabilistic measure for how deep SW ions penetrate into the atmosphere before becoming nascent ENAs. 
Figure \ref{ENA_Prod} shows how few nascent ENAs are created above 300 km and below 150 km there are no longer any SW ions present as they have all become nascent ENAs. 
It is worth noting that the minimum {\color{black}SA} atmosphere model has the lowest number density for atmospheric species at a given altitude and thus has ENA production occurring {\color{black}deeper} in the atmosphere than the mean or maximum SA atmosphere models. 
Also, for all three atmosphere models, nascent helium ENA production occurred deeper in the atmosphere than nascent hydrogen ENA production which may be attributed to smaller average charge exchange cross sections for helium ions. 

The following sections describe how the nascent ENA energy and altitude distributions detailed above, as well as the cross sections outlined in the previous section, are used to transport large ensembles of ENA particles in the atmosphere of Mars using MC methods. 

\section{Monte Carlo Transport of Precipitating ENAs}

The process of energy-momentum transfer and thermalization is inherently stochastic as a multitude of probabilistic events occur to bring the system from the initial state (energetic) to the final state (thermal).
To model such systems, MC methods may be employed with detailed knowledge of atomic and molecular interactions, to accumulate macro statistics of the system.   
In this study, large ensembles of particles are modeled using MC simulations to obtain probabilities for macro events such as atmospheric heating rates, secondary hot atom production, and atmospheric escape fluxes. 
A high quality, parallel random number generator was employed to ensure the random numbers used for stochastic events lack any pattern \citep[]{Ladd:2008}. 

The general scheme of the MC simulation includes an ensemble of ENA test particles which are transported through the atmosphere of Mars as they collide with neutral target atoms and molecules until the test particles either thermalize with the atmosphere or, for upward moving particles, reach a conditional collisionless upper boundary altitude of 700 km with an energy above escape energy. 
The test particles in the simulation are allowed to move in {\color{black}3D}, although the atmosphere models used are essentially {\color{black}1D}, only considering altitude in determining neutral densities. 

The simulation is initialized by determining random ENA energies and starting altitudes according to cumulative probabilities for nascent ENA production described in the previous section while the x-y coordinates are initialized to zero. 
All test particle ensembles are investigated with a {\color{black}solar zenith angle} of 0 deg in this work so the initial unit velocity vectors are directly incident on the surface of Mars.  
The case of non-zero {\color{black}solar zenith angle} can easily be computed without modification of the considered MC methods. 

After the initial conditions are set for the test particles, they are transported through the atmosphere using a step-by-step method. 
At each step, the total mean free path is determined for each test particle at it's current altitude
\begin{equation}
	\lambda(z,E) = \frac{1}{ \displaystyle\sum_{i} n_{i}(z) \sigma_{i}(E)}
\end{equation}
where $n_{i}(z)$ is the density of the $i^{th}$ atmospheric species at altitude $z$ and $\sigma_{i}(E)$ is the collisional cross section between the ENA test particle and the $i^{th}$ atmospheric species. 
The step-size employed at a given location in the atmosphere is determined using this total mean free path. 
Following methods used by \cite{Fox:2009}, the step-size, $\Delta s$, utilized is either 20\% of the total mean free path, $\lambda(z,E)$, or 1 km, which ever value is smaller.
This methodology ensures that the atmospheric densities are near constant along the step-size $\Delta s$, before and after the test particle is transported. 

With the total mean free path and step-size determined, the probability for a collision to occur within that step-size is defined by
\begin{equation}
	P_{coll} = 1 - \exp \left[ \frac{-\Delta s}{\lambda(z,E)} \right]
	\label{Prob_Coll}
\end{equation}  
which is a simplified expression of typical particle transport, determined in utilizing straight trajectory transport over these small step-sizes which is appropriate for keV/amu energies.
A random number, $\eta$, determines if a collision occurs in the interval $\Delta s$.
If a collision does occur, the exact location of the collision, $\Delta l$, may be found using Equation \ref{Prob_Coll} with $\eta = P_{coll}$ and solving for $\Delta s = \Delta l$, found to be
\begin{equation}
	\Delta l = - \lambda(z,E) \log \eta
\end{equation}
where it is always the case that the distance to the collision location $\Delta l$ is less than the step-size $\Delta s$. 
For a given collision, the target species is determined using effective atmospheric mixing ratios such that
\begin{equation}
	\Gamma_{i}(z) = \frac{n_{i}(z) \sigma_{i}(E)}{\displaystyle\sum_{j} n_{j}(z) \sigma_{j}(E)}
\end{equation}
where $\displaystyle\sum_{i} \Gamma_{i}(z) = 1$. 
An array was then formed using the calculated mixing ratios and a random number was selected to determine the target species. 
The scattering angle for a given collision was determined, as discussed in Section 2, with azimuthal scattering angles being randomly generated between 0 and 2$\pi$.
With scattering angles $\theta$ and $\phi$, the normalized directional cosines for the test particle velocity may be updated such that
\begin{equation}
\begin{split}
	u_{x}' &= \frac{\sin \theta}{\alpha} \left[ u_{x} u_{z} \cos \phi - u_{y} \sin \phi \right] + u_{x} \cos \theta, \\
	u_{y}' &= \frac{\sin \theta}{\alpha} \left[ u_{y} u_{z} \cos \phi + u_{x} \sin \phi \right] + u_{y} \cos \theta, \\
	u_{z}' &= -\alpha \sin \theta \cos \phi + u_{z} \cos \theta,
\end{split}
\label{dir_cos}
\end{equation} 
where $\alpha = \sqrt{ 1 - u_{z}^2}$ and the vectors $\vec{u}$ and $\vec{u}'$ are the directional cosines before and after the collision. 
In the instance of $u_{z} = \pm 1$, Equation \ref{dir_cos} reduces to 
\begin{equation}
\begin{split}
	u_{x}' &= \sin \theta \cos \phi \\
	u_{y}' &= \pm \sin \theta \sin \phi \\
	u_{z}' &= \pm \cos \theta
\end{split}
\end{equation}
which occurs during the first collision if the initial {\color{black}solar zenith angle} is 0 deg.
In the trivial case of no collision occurring for a given step, $\vec{u}' = \vec{u}$.  

The updated particle location, $\vec{r}'$, is found as either the location of the previous collision, or in the case of no collision in the previous step, a distance $\Delta s$ from the previous step's position $\vec{r}$. 
Using the previous step's directional cosine, $\vec{u}$ 
\begin{equation}
	\vec{r}' = \vec{r} + \Delta \vec{u}
\end{equation}
where $\Delta=\Delta l$ if a collision occurs and $\Delta=\Delta s$ if no collision occurs. 

Following a collision, the energy lost by the fast test particle is calculated in the lab frame
\begin{equation}
	\delta \epsilon(\epsilon, \theta) = \epsilon \frac{ 2 m_{p} m_{t} }{(m_{p} + m_{t})^2} ( 1 - \cos \theta )
	\label{dE}
\end{equation}
using the lab frame energy of the test particle $\epsilon$, the {\color{black}center of mass} frame scattering angle $\theta$, and the masses of the projectile and target particles, $m_{p}$ and $m_{t}$. 
In this study, the keV energy of the ENAs is so much larger than that of the thermal atmospheric energy that to describe collisions we treat all target particles as being motionless in the atmosphere.
Using the motionless atmosphere approximation as well as conservation of energy, the energy loss from Equation \ref{dE} is also the energy transferred to the target particle. 
If the energy transferred to the target particle is significantly larger than the thermal energy of the atmosphere, the recoil target atoms and molecules are considered SH atoms/molecules. 

Using the MC transport algorithm described above, statistics were obtained for energy relaxation of precipitating ENAs as well as SH atom/molecule production. 
Details of ENA propagation through the planetary atmosphere and reflection from atmospheric layers, time-dependent energy distributions, and thermalization properties are outlined in the following sections.  

\vspace{10pt}
\section{Results of Monte Carlo Simulations for SW Ion and ENA Precipitation}

Several key parameters were extracted from the MC simulations to best display physical details of the thermalization and possible multi-collision backscattering of incident ENAs as well as the production of upward and escape fluxes of SH atoms/molecules in the atmosphere.  
Results are presented for realistic initial energy-altitude distributions of the SW ions and produced nascent ENAs using the different atmospheric density models for minimum, mean, and maximum SA, all of which is described in Section 3. 
In addition, results for mono-energetic ensembles of hydrogen atoms with an initial energy of 1 keV, and helium atoms with initial energy of 4 keV, are also included for comparison. 
The energies of the mono-energetic ensembles were chosen as they represent the most common energies associated with the model SW energy distribution. 
These mono-energetic ensembles of nascent ENAs, which are introduced for illustrative purposes, were all initialized at an altitude of 200 km, which is the average altitude of nascent ENA production as seen in Figure \ref{ENA_Prod}, and were transported using the mean SA neutral atmosphere model. 
Ensembles for hydrogen and helium ENAs whose collisions are described by the HS cross sections, were also considered to compare with the anisotropic, QM cross sections. 
HS cross sections have been obtained from Van der Waals radii used for the physical radii of all atoms and molecules \citep[]{Bondi:1964}. 
These HS ensembles were initialized using the realistic energy-altitude ENA distributions and were transported through the mean SA neutral model atmosphere. 
All ensembles listed above contained 50,000 test particles which was double the number of test particles required for saturated statistics of all processes of interest. 

Results were obtained from the simulations using discrete, 2D probability densities $f(x_{i},y_{j})$, for two parameters of interest, $x_{i}$ and $y_{j}$, where each test particle was placed in the appropriate discrete bin at every transport step.
The probability densities are normalized such that
\begin{equation}
\displaystyle\sum_{i,j} f(x_{i},y_{j}) \, \Delta x \Delta y = 1, 
\end{equation} 
where $\Delta x$ and $\Delta y$ are the bin sizes for the $x$ and $y$ parameters. 
Average values for a given parameter were obtained using a weighted average
\begin{equation}
<y_{j}> = \frac{\displaystyle\sum_{i} x_{i} f(x_{i},y_{j})}{\displaystyle\sum_{i} f(x_{i},y_{j})}, 
\label{AvgProb}
\end{equation}
where $<y_{j}>$ is the weighted average value.

Details of the results extracted from the MC simulations follow. 


\subsection{Thermalization}
We classify an ENA test particle as being ``thermalized'' in this work if all of the original energy of a few keV/amu is transfered to other atmospheric particles during the simulation, thus making the test particle thermal with its environment.
A formal cutoff energy of 0.1 eV was used to determine if the test particles had thermalized. 
This energy was chosen as it is slightly above the mean thermal energy of 0.02 eV at a typical temperature of 200 K found in the upper atmosphere of Mars \citep[]{Krasnopolsky:2002}. 

\begin{figure*}
\centering
\vspace{25pt}
\includegraphics[width=.35\linewidth]{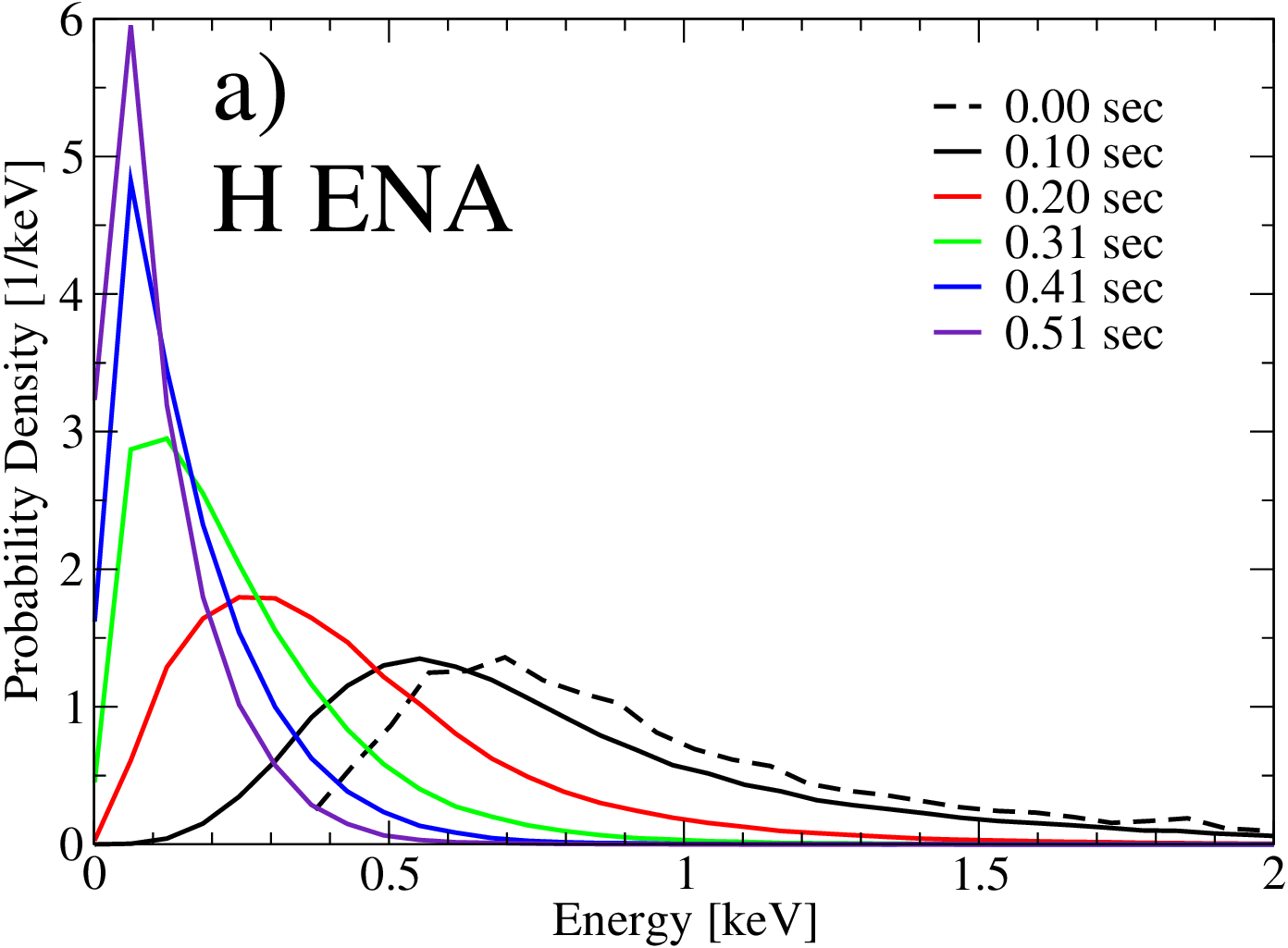}
\hspace{10pt}
\includegraphics[width=.35\linewidth]{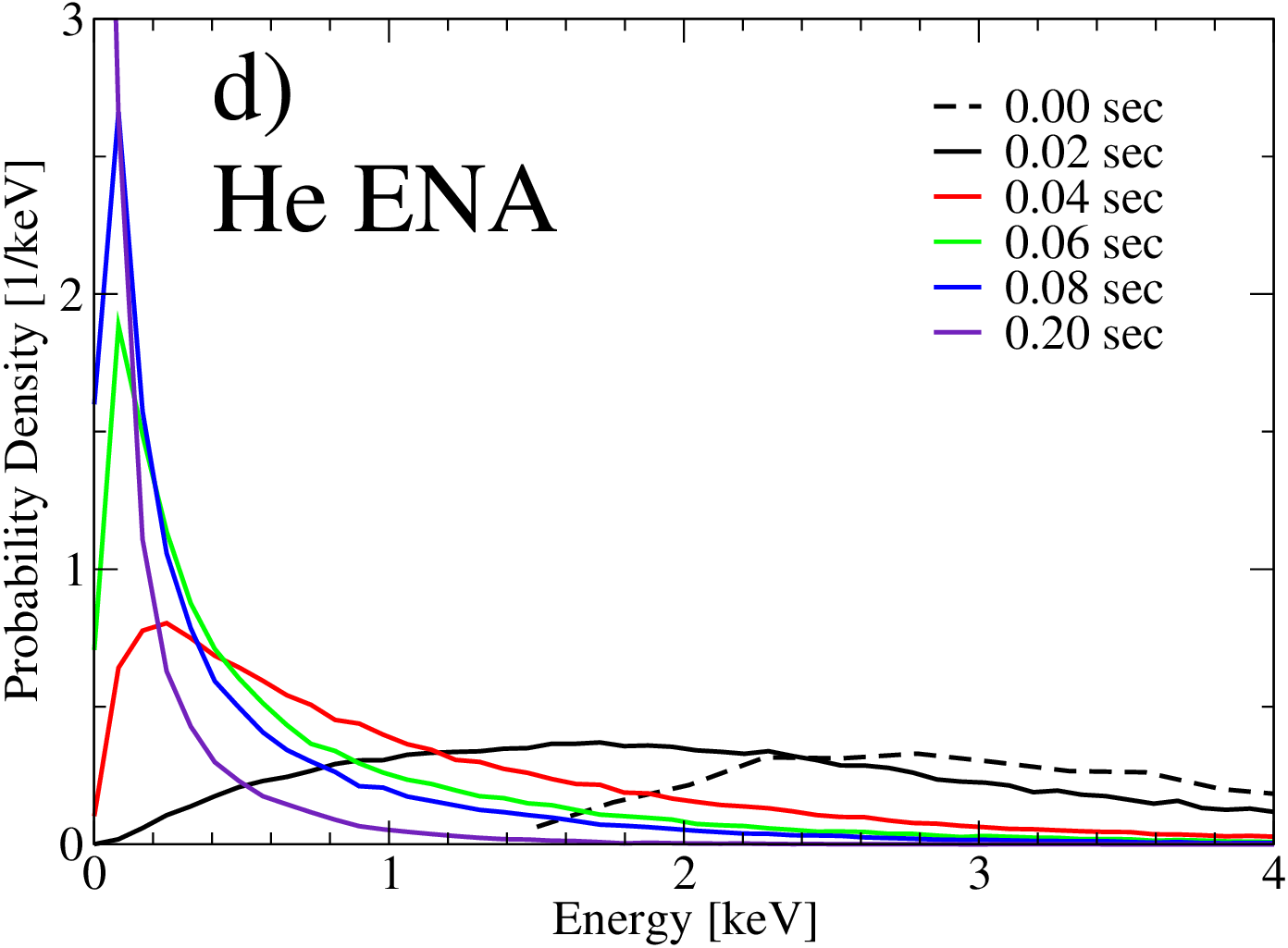} \\
\vspace{25pt}
\includegraphics[width=.35\linewidth]{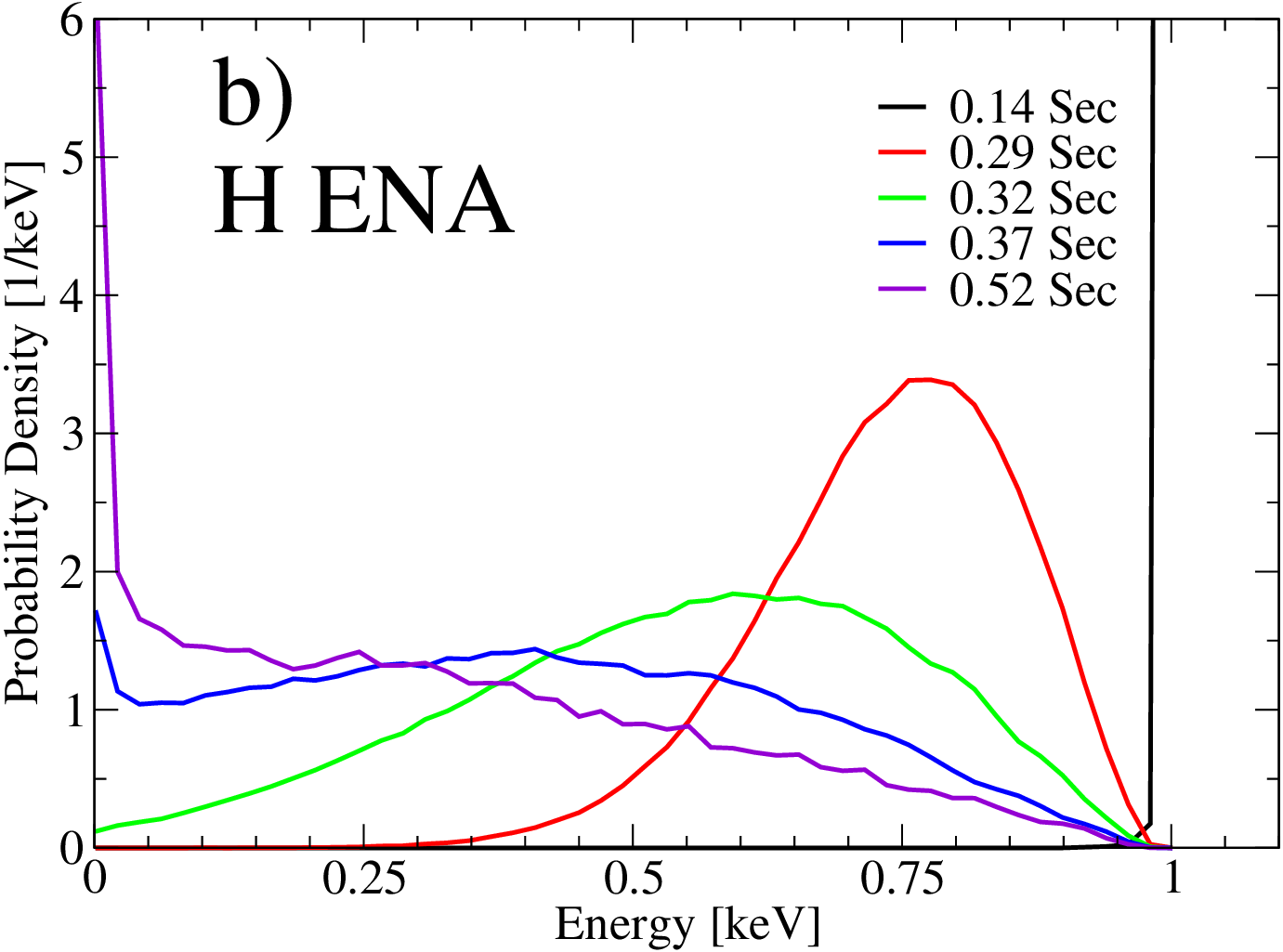}
\hspace{10pt}
\includegraphics[width=.35\linewidth]{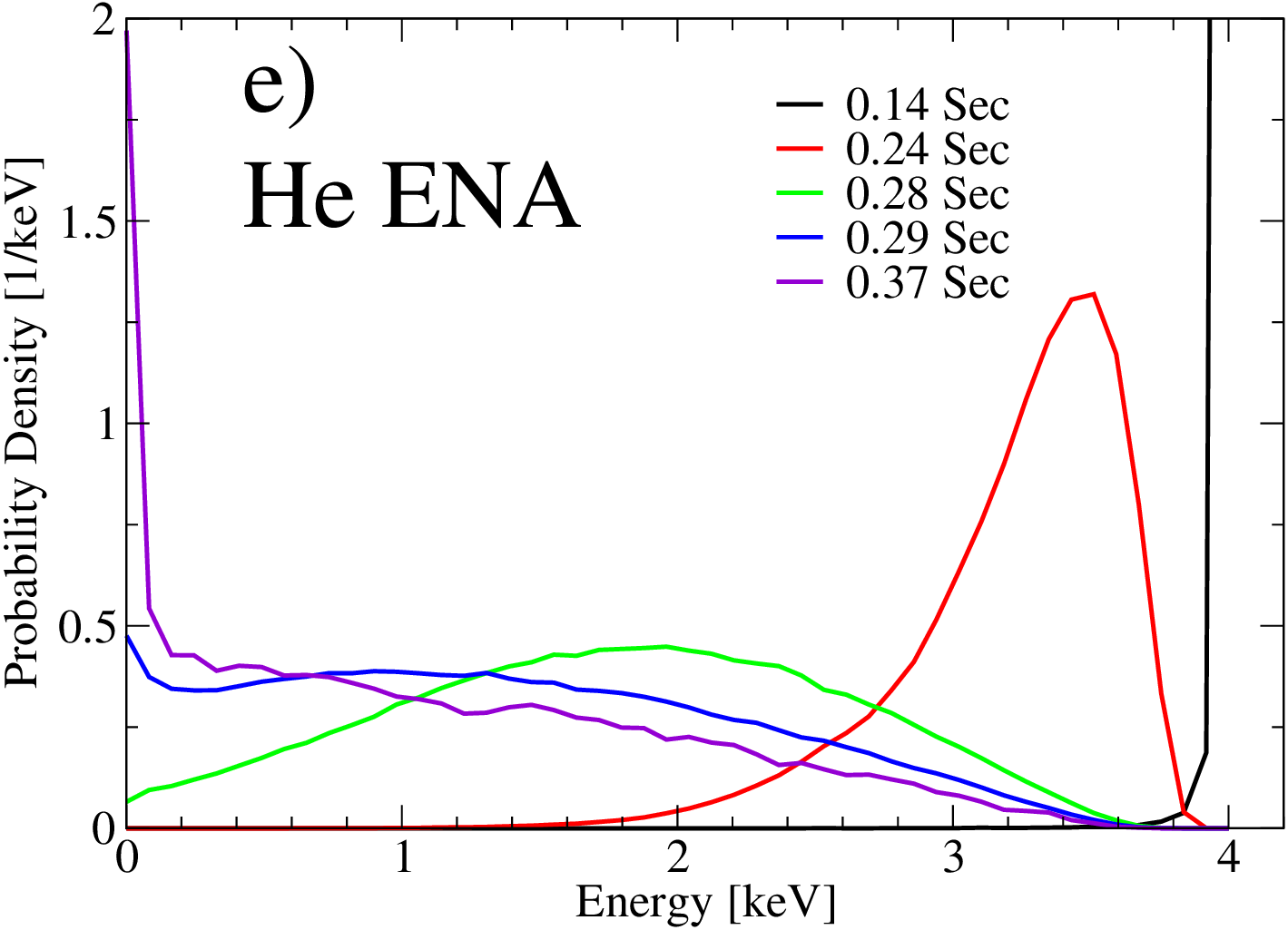} \\
\vspace{25pt}
\includegraphics[width=.35\linewidth]{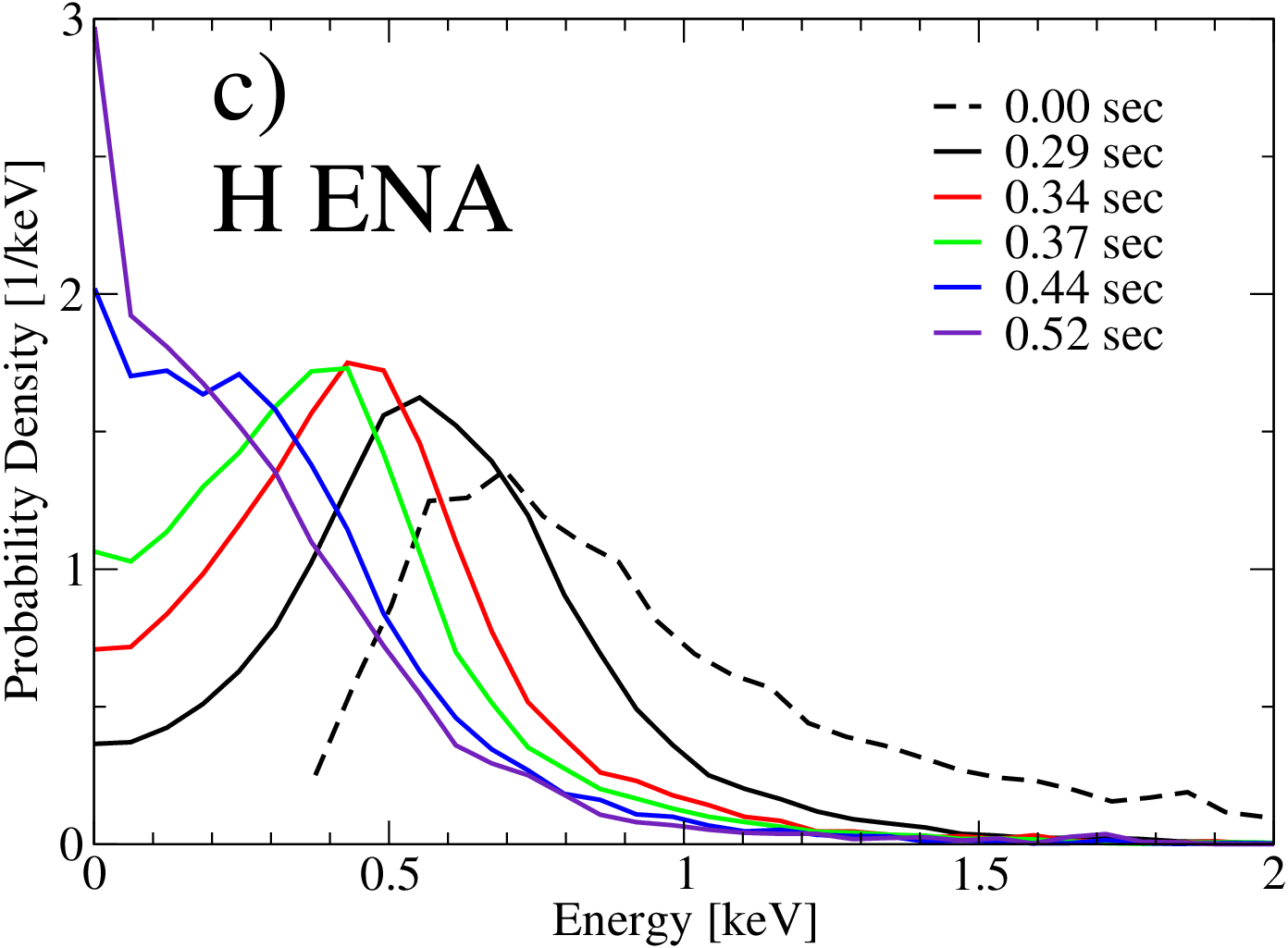}
\hspace{10pt}
\includegraphics[width=.35\linewidth]{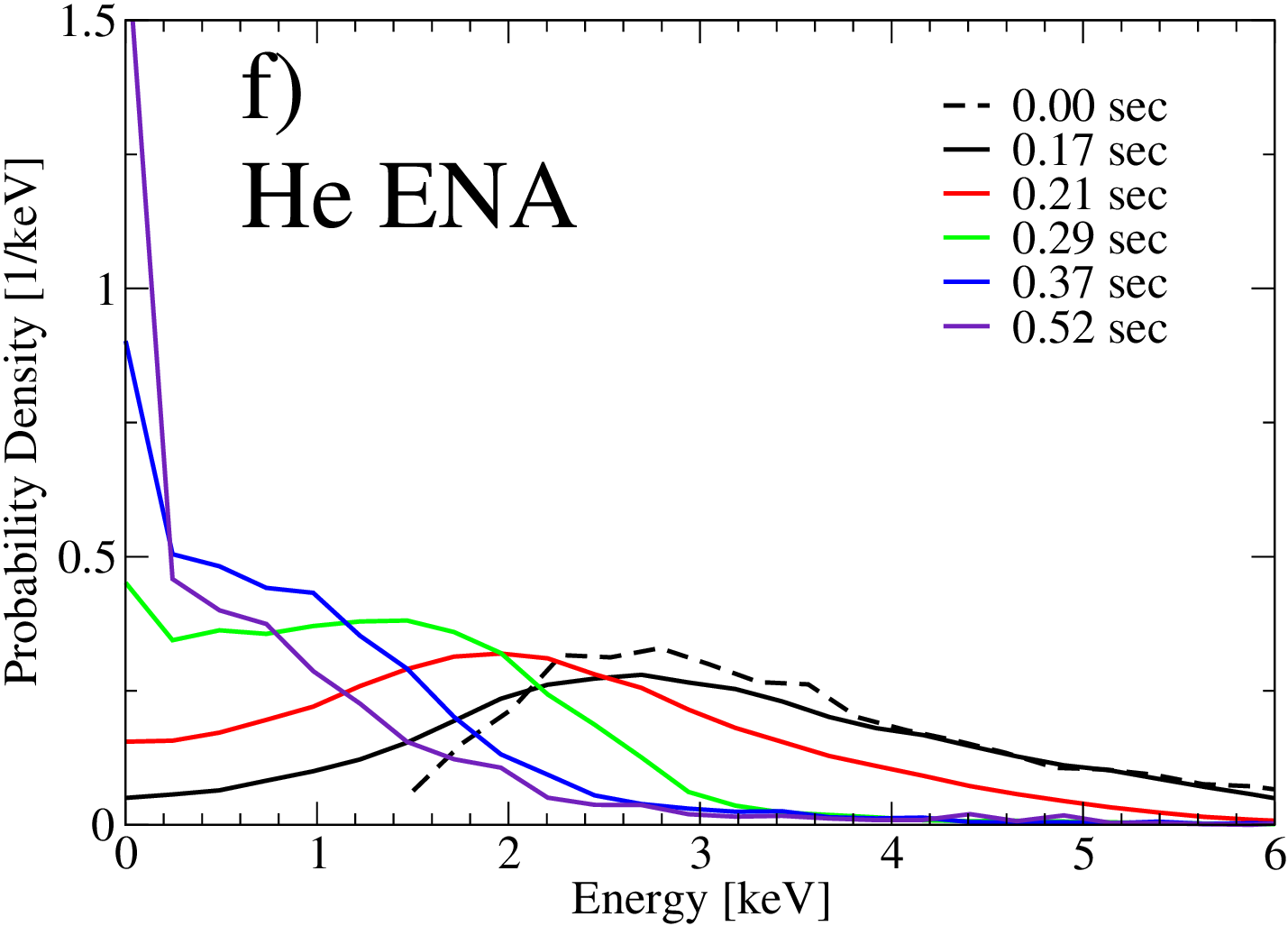} \\
\caption{\footnotesize Energy distributions for precipitating hydrogen ENAs are shown for HS cross sections a), mono-energetic 1 keV b), and realistic SW energy c) ensembles, all utilizing the mean SA model atmosphere. Results are shown in a similar manner for precipitating helium ENAs in d--f) with the mono-energetic ensemble having an energy of 4 keV. Initial, $t$=0 sec, energy distributions are shown for ensembles utilizing realistic SW energy distributions.}
\label{TimeSlices}
\end{figure*}

Analysis of energy distributions of precipitating ENAs are extremely useful as they provide insight into the time-dependent thermalization process of the ensembles. 
Time-dependent energy distributions of precipitating particles allow the evaluation of rates involved with ENA energy relaxation and atmospheric heating for different parameters of precipitating fluxes and upper atmosphere conditions of Mars. 
Figure \ref{TimeSlices} displays energy distributions for the HS, mono-energetic, and realistic SW energy ensembles for both hydrogen and helium ENAs, all of which utilizing the mean SA neutral atmosphere model. 
The HS ensembles shown in Figure{\color{black}s} \ref{TimeSlices}{\color{black}a) and \ref{TimeSlices}d)} lose the vast majority of their energy very quickly yet take more time to completely thermalize than the realistic SW or mono-energetic ensembles, both of which utilize QM cross sections.
The mono-energetic and realistic SW ensembles shown in Figure{\color{black}s} \ref{TimeSlices}{\color{black}b) and \ref{TimeSlices}e) and Figures \ref{TimeSlices}c) and \ref{TimeSlices}f)} display how the helium ENAs lose their energy much slower than the hydrogen ENAs.  
In both of the realistic SW energy distribution ensembles, an interesting feature can be seen at times 0.29, 0.37, and 0.52 seconds which resembles the mixing of two fluxes with different speeds as the lower {\color{black}energy} portion of the ensembles {\color{black}thermalize} first, leading to a low energy peak in the distributions, followed by a higher energy peak which slowly {\color{black}merges} with the first peak. 
This feature is not seen in either the HS or mono-energetic ensembles as it is a result of both small angle, anisotropic QM cross sections used in these ensembles as well as initial energy distributions with a large spread in energies.  

\begin{figure}
	\vspace{20pt}
	\includegraphics[width=.85\linewidth]{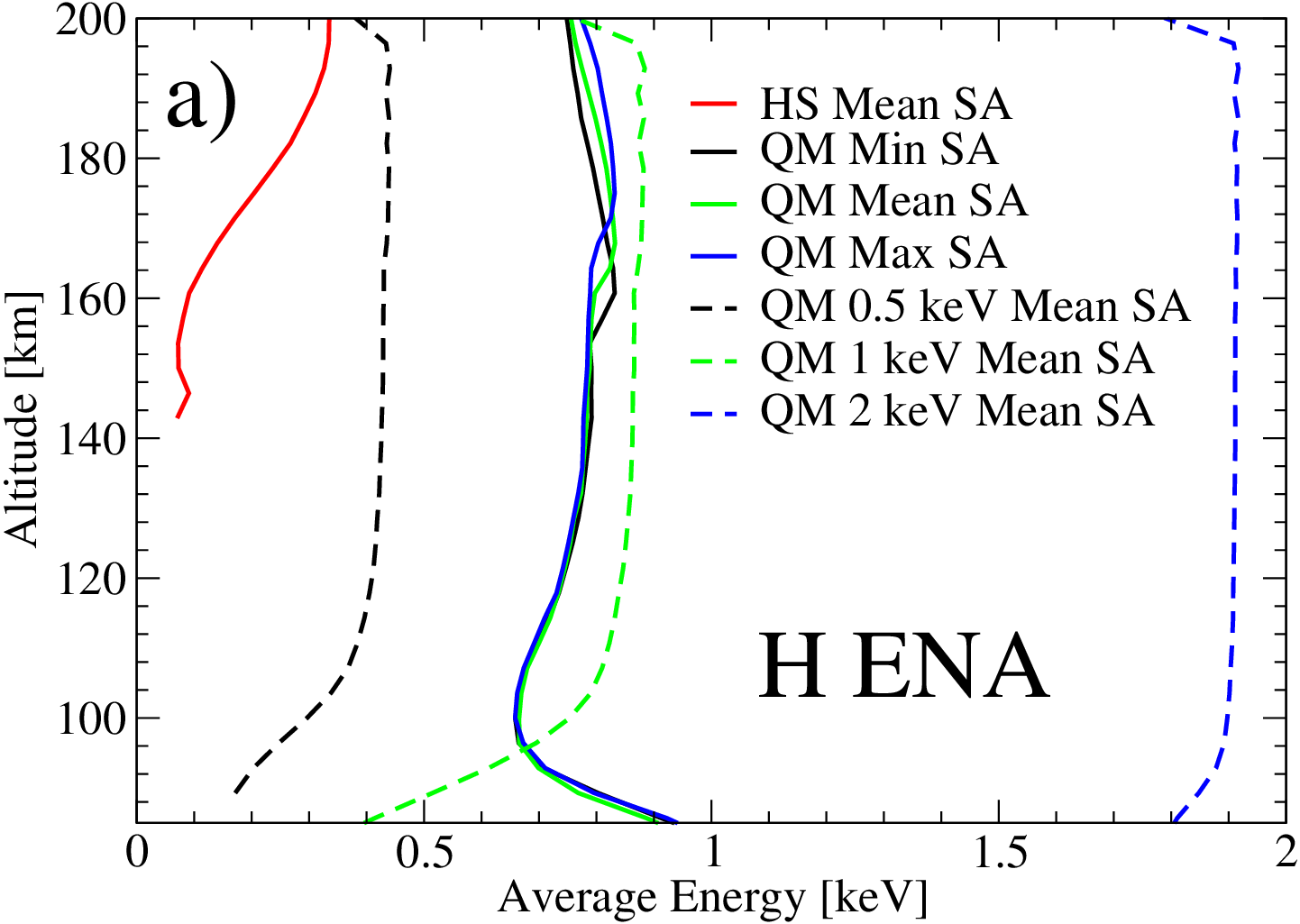}\hfill
	\vspace{30pt}
	\includegraphics[width=.85\linewidth]{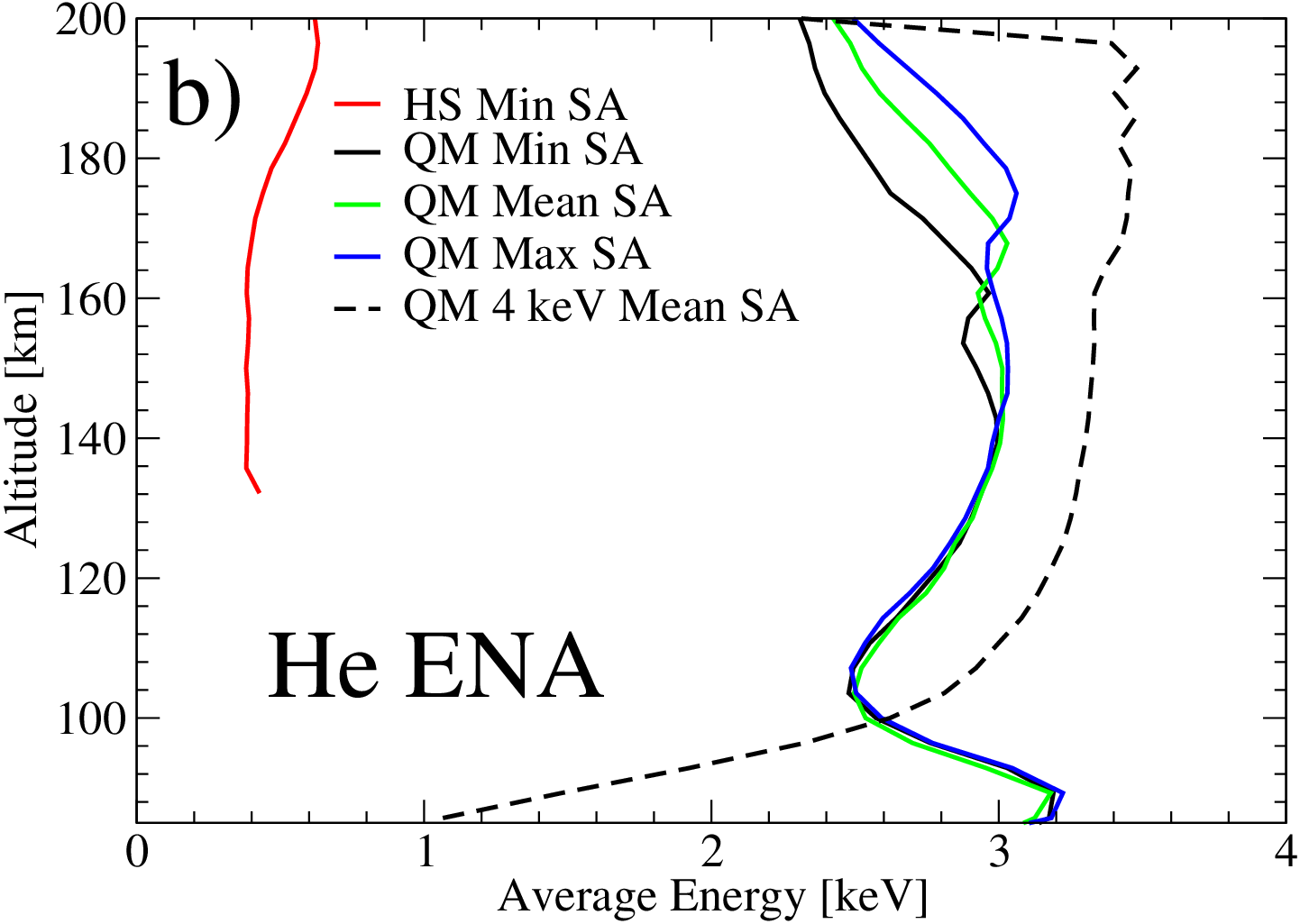} \\
	\caption{\footnotesize Average weighted energy for both hydrogen a) and helium b) ENAs as a function of altitude for all three atmosphere models. Additionally, ensembles utilizing HS cross sections and mono-energetic ensembles are also shown for comparison.}
	\label{Avg_engy}
\end{figure}

Figure \ref{Avg_engy} displays the average energy of the test particle ensembles as a function of altitude in the Mars atmosphere for hydrogen and helium ENAs utilizing Equation \ref{AvgProb}. 
An unexpected feature can be seen in Figure \ref{Avg_engy} in that there is a steady decrease in average energy with decreasing altitude starting from 160 km as the ENAs penetrate into the atmosphere, yet there is an altitude for both hydrogen and helium ENAs at which the average energy begins to increase. 
Additional mono-energetic ensembles for hydrogen ENAs with energies of 0.5 and 2 keV are shown in Figure \ref{Avg_engy}a to better understand the average energy increases for the realistic SW ensembles. 
These increases do not exist for mono-energetic ensembles which decrease monotonically with decreasing altitude and penetrate deeper into the atmosphere as the initial energy is increased. 
The altitude, at which the average energy begins to increase, is located at 100 km and 110 km for hydrogen and helium respectively and is attributed to the average penetration depths of the ENAs with realistic SW energies. 
This increase in average energy, along with a significant decrease in numbers of precipitating particles, can then be attributed to the fact that these low altitudes are only accessible to the most energetic portion of the initial ENA energy distribution and thus the total average energy of the ensemble jumps from including all particles to including only the most energetic ones. 
The relative energy losses become smaller as the collision energy increases thus making a fraction of the energetic particles more abundant at low altitudes. 
Figure \ref{Avg_engy} also displays the differences in average energy due to different atmospheric neutral density models. 
These differences are extremely small for altitudes less than 160 km yet begin to play a larger role for higher altitudes.
For example, average energies differ by 50 eV and 200 eV between min and max SA atmosphere models at an altitude of 200 km for hydrogen and helium ENAs respectively. 
In comparison to realistic QM cross sections, ensembles utilizing isotropic HS cross sections have large energy losses observed at higher altitudes, greater than 200 km, as seen in Figure \ref{Avg_engy}, resulting in a total penetration depth of 140 km and 130 km for hydrogen and helium ENAs, both of which are much lower then their QM counterparts. 

\begin{figure}
	\vspace{20pt}
	\includegraphics[width=.85\linewidth]{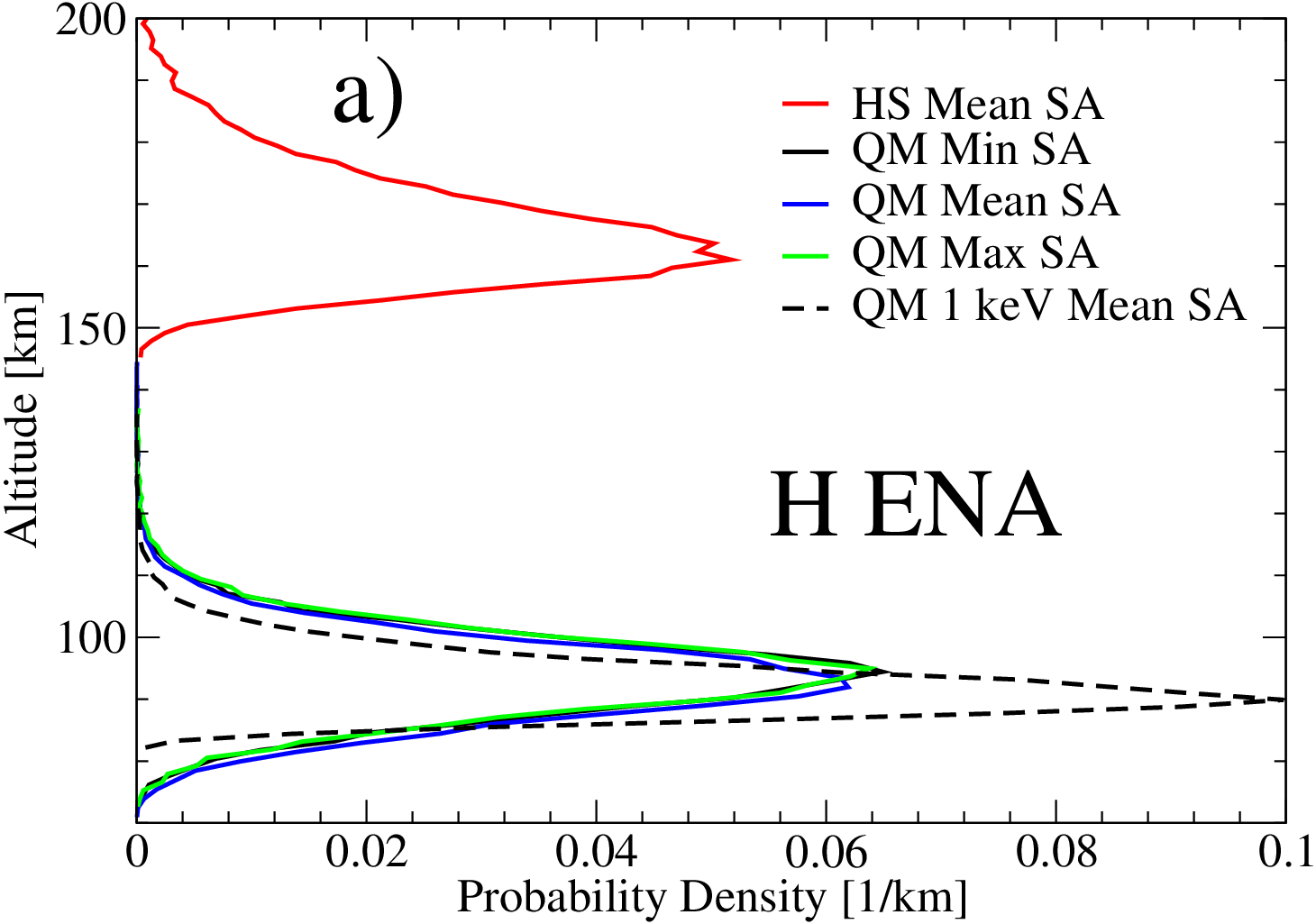}\hfill
	\vspace{30pt}
	\includegraphics[width=.85\linewidth]{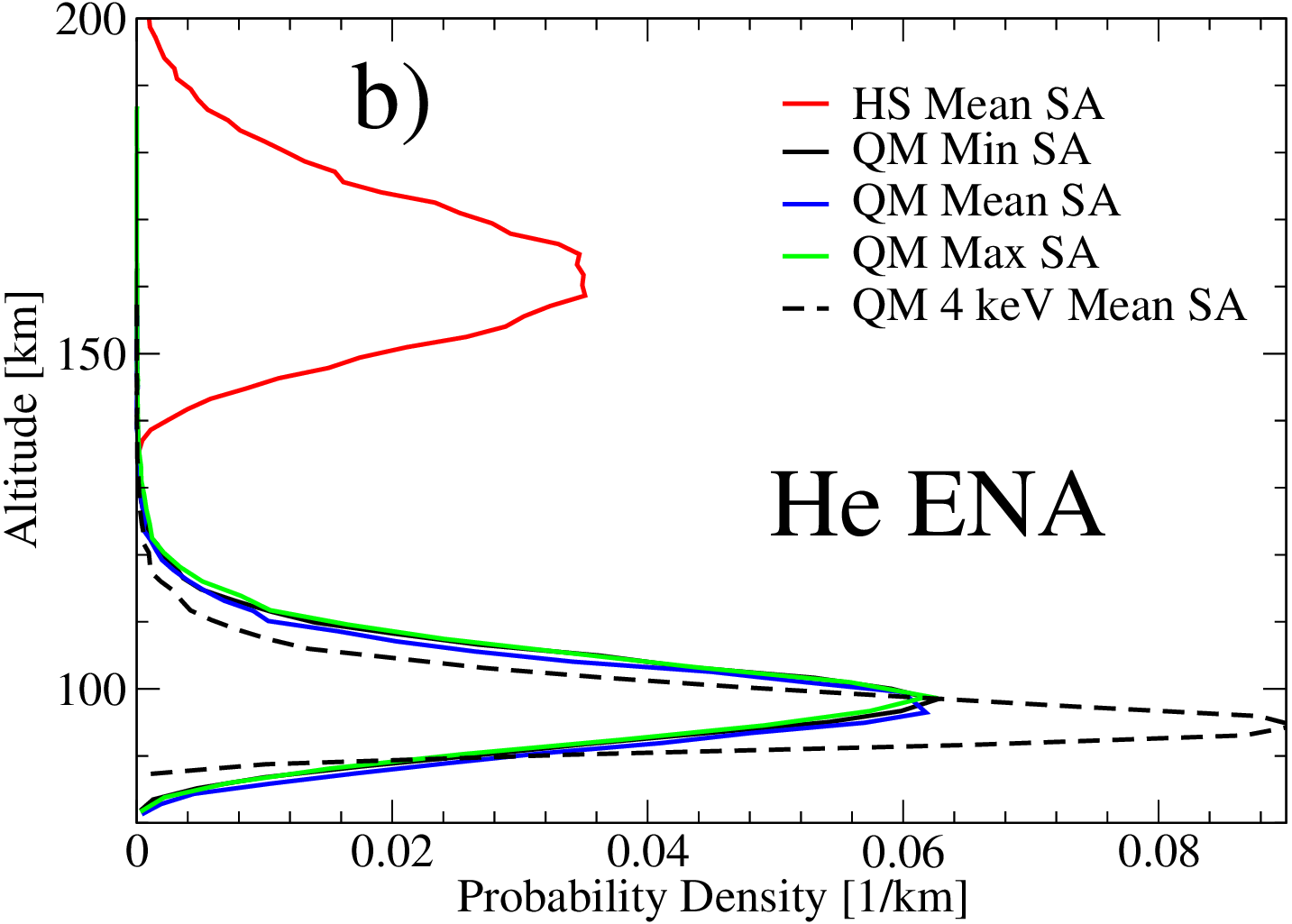} \\
	\caption{Thermalization altitude probability densities for both hydrogen a) and helium b) ENAs using all three atmosphere models. Additionally, ensembles utilizing HS cross sections and mono-energetic ensembles are also shown for comparison.}   
	\label{Therm_Alt}
\end{figure}

The altitude at which test particles thermalize with the atmosphere was also obtained from the MC simulations for all ensembles.
Figure \ref{Therm_Alt} displays probability densities for the thermalization altitude of all ensembles. 
Very little difference in thermalization altitude can be seen in comparing the three different atmosphere models with all three models having the highest thermalization probability at 95 km and 98 km for hydrogen and helium respectively. 
Drastic differences between the HS ensembles and the QM ensembles can be seen in Figure \ref{Therm_Alt} with the highest thermalization probability occurring at 160 km for both hydrogen and helium HS ensembles. 
The probability distribution for the helium ensemble using HS cross sections is wider then its hydrogen counterpart, being due to higher rate of collision for the helium ENAs with larger HS radii \citep[]{Bondi:1964}.  
The mono-energetic distributions have slightly deeper thermalization depths and narrower distributions than the SW energy distribution ensembles, demonstrating the collective effects of having an initial energy distribution as compared with {\color{black}a} single initial energy. 

\begin{figure}
	\vspace{20pt}
	\includegraphics[width=.85\linewidth]{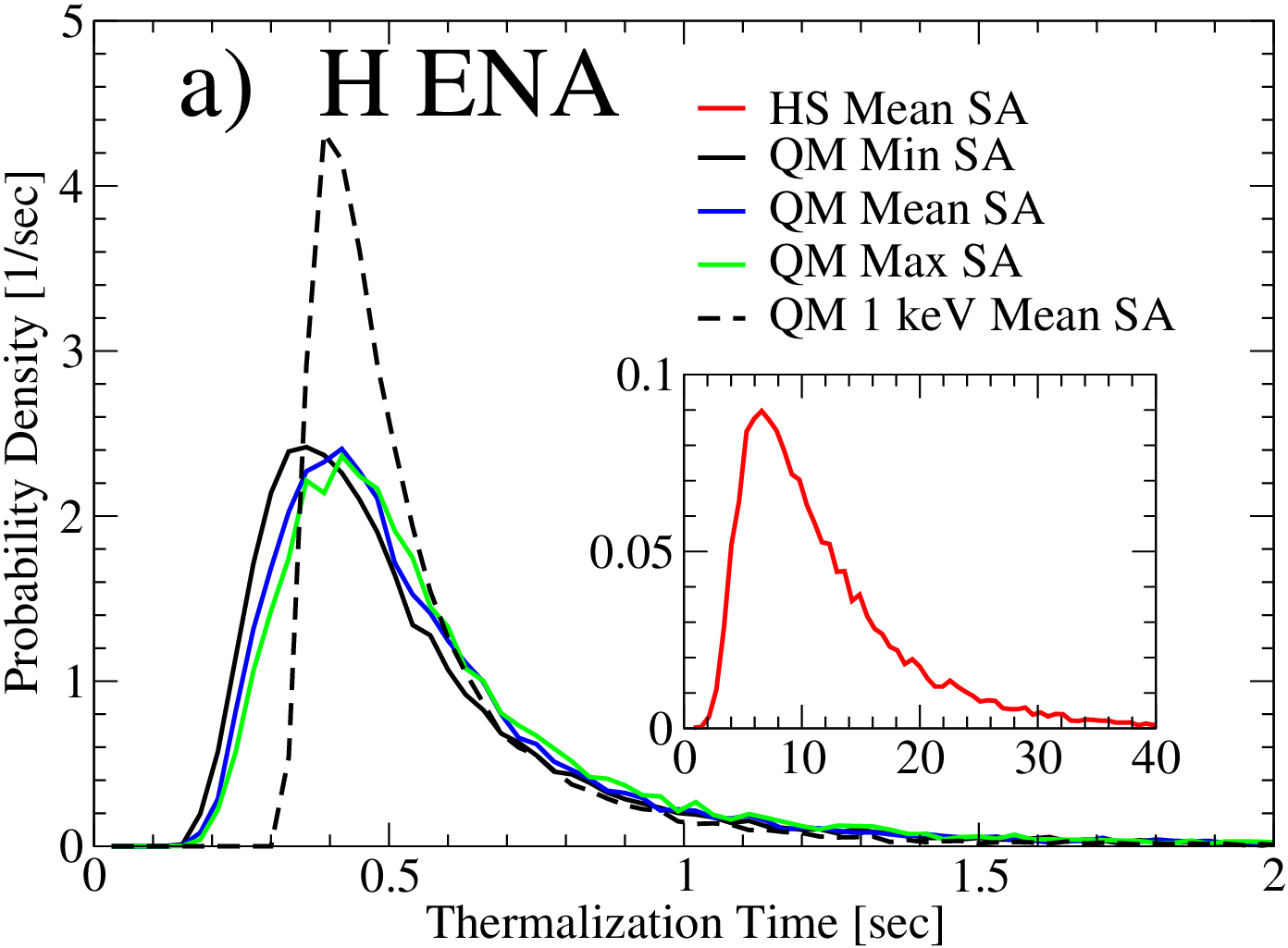}\hfill
	\vspace{30pt}
	\includegraphics[width=.85\linewidth]{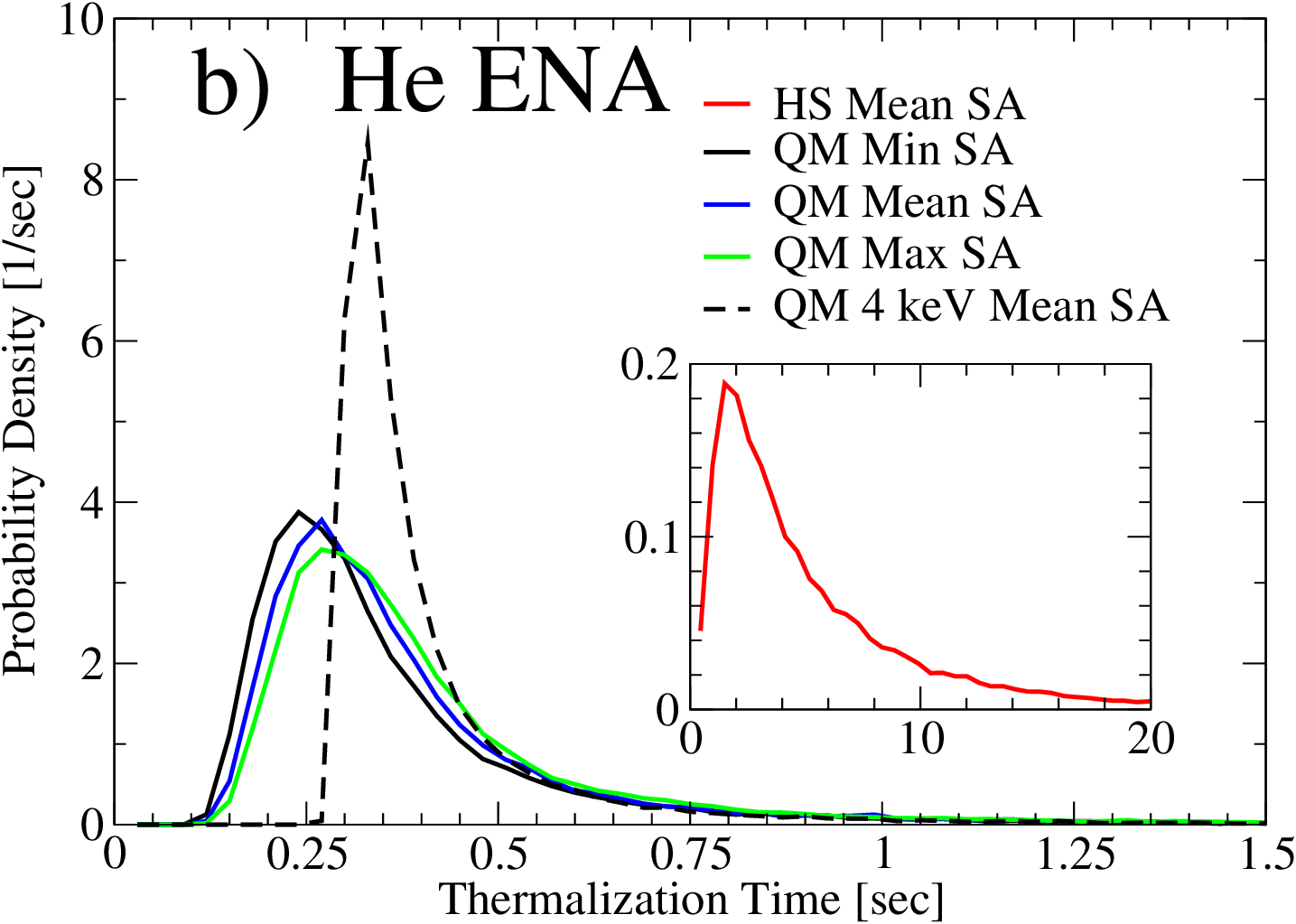} \\
	\caption{\footnotesize Thermalization time probability densities for both hydrogen a) and helium b) ENAs using all three atmosphere models as well as ensembles utilizing HS cross sections and mono-energetic ensembles.}   
	\label{Therm_Time}
\end{figure}

In addition to thermalization altitudes, thermalization times were also obtained from the MC simulations for all ensembles. 
Thermalization time probability densities for all ensembles are shown in Figure \ref{Therm_Time}.
Slightly larger average scattering angles for a given ENA velocity, shown in Figure \ref{Rand_Angles}, leads to faster thermalization of helium ENAs as compared to hydrogen ENAs, an effect which is clear from Figure \ref{Therm_Time}. 
The mono-energetic probability distributions are narrower than both of their respective realistic SW energy distribution ensembles, yet the tails of all ensemble distributions decay in a very similar manner starting at 0.5 seconds. 
The thermalization time probability density for HS cross sections in Figure \ref{Therm_Time} is vastly different from the QM ensembles, with average thermalzation times an order of magnitude larger than their QM counterparts. 
These large thermalization times are attributed to the test particles in the HS ensembles occupying only high altitude layers of the atmosphere, above 150 km, as seen in Figures \ref{Avg_engy} and \ref{Therm_Alt}. 
The high altitude atmosphere layers have significantly smaller neutral densities and thus a lower frequency of thermalizing collisions as compared with lower atmosphere layers where the ensembles of test particles utilizing QM cross sections thermalize.
For example, the neutral density increases by 3--4 orders of magnitude from 150 km to 100 km \citep[]{Krasnopolsky:2002}. 
The average mean free path for HS ensembles within the layer of atmosphere between 150--200 km, where the majority of ENAs using HS cross sections thermalize, ranges from several to tens of kms. 
These large mean free paths lead to long transport times between collisions, in particular, ENAs with energies less than 20 eV are transported for 0.1 seconds or more on average.
The combination of these long transport times in between collisions and hundreds of collisions before thermalization results in the longer thermalization times for HS collisions, shown in Figure \ref{Therm_Time}. 


\subsection{Secondary Hot Atoms and Molecules}
During the transport and thermalization of the ENA test particles, several thousand collisions occur with the atmospheric gases, the vast majority of which happen with very small scattering angles, thus transferring to the ambient gas a very small amount of energy as determined by Equation \ref{dE}. 
With so many collisions occurring during the thermalization process, several low probability, large-angle scattering events transpire during the lifetime of a test particle which provide enough energy to the target particle for it to be considered hot itself. 
The energy transfer threshold to be considered a SH atom/molecule was set to 0.1 eV which is significantly higher than atmospheric thermal energies. 

\begin{figure*}
\centering
	\vspace{20pt}
\includegraphics[width=.35\linewidth]{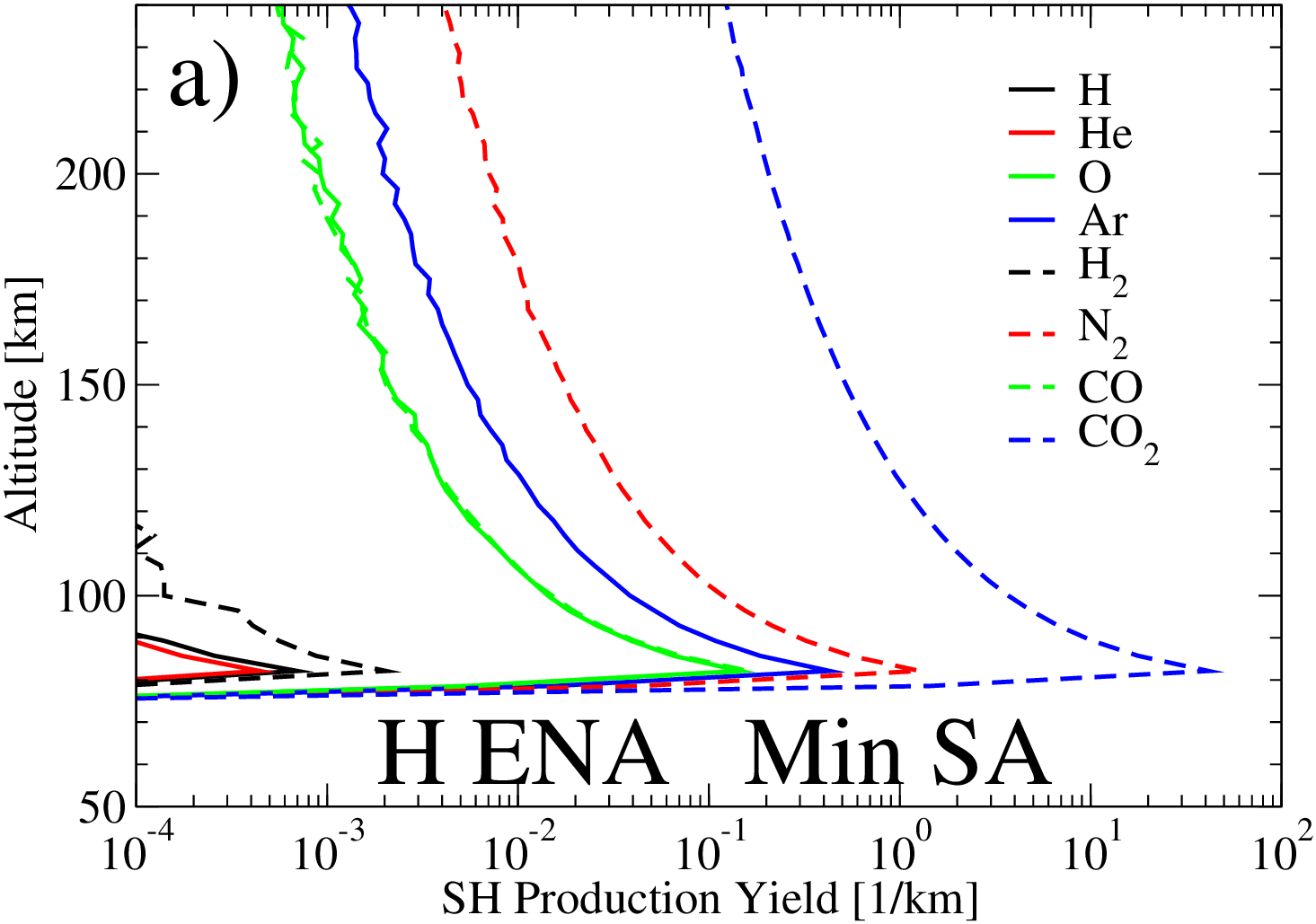}
\includegraphics[width=.35\linewidth]{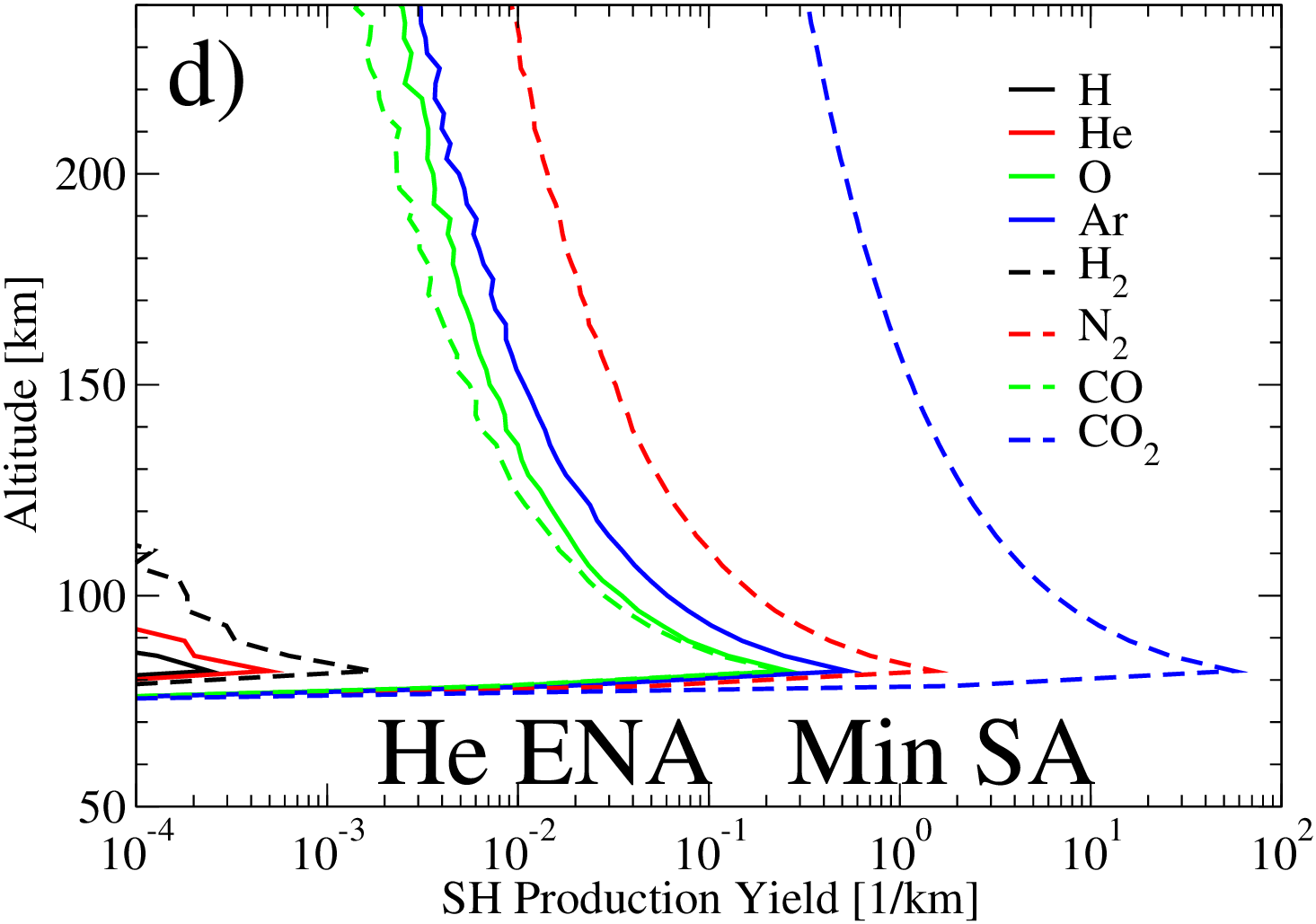} \\
	\vspace{25pt}
\includegraphics[width=.35\linewidth]{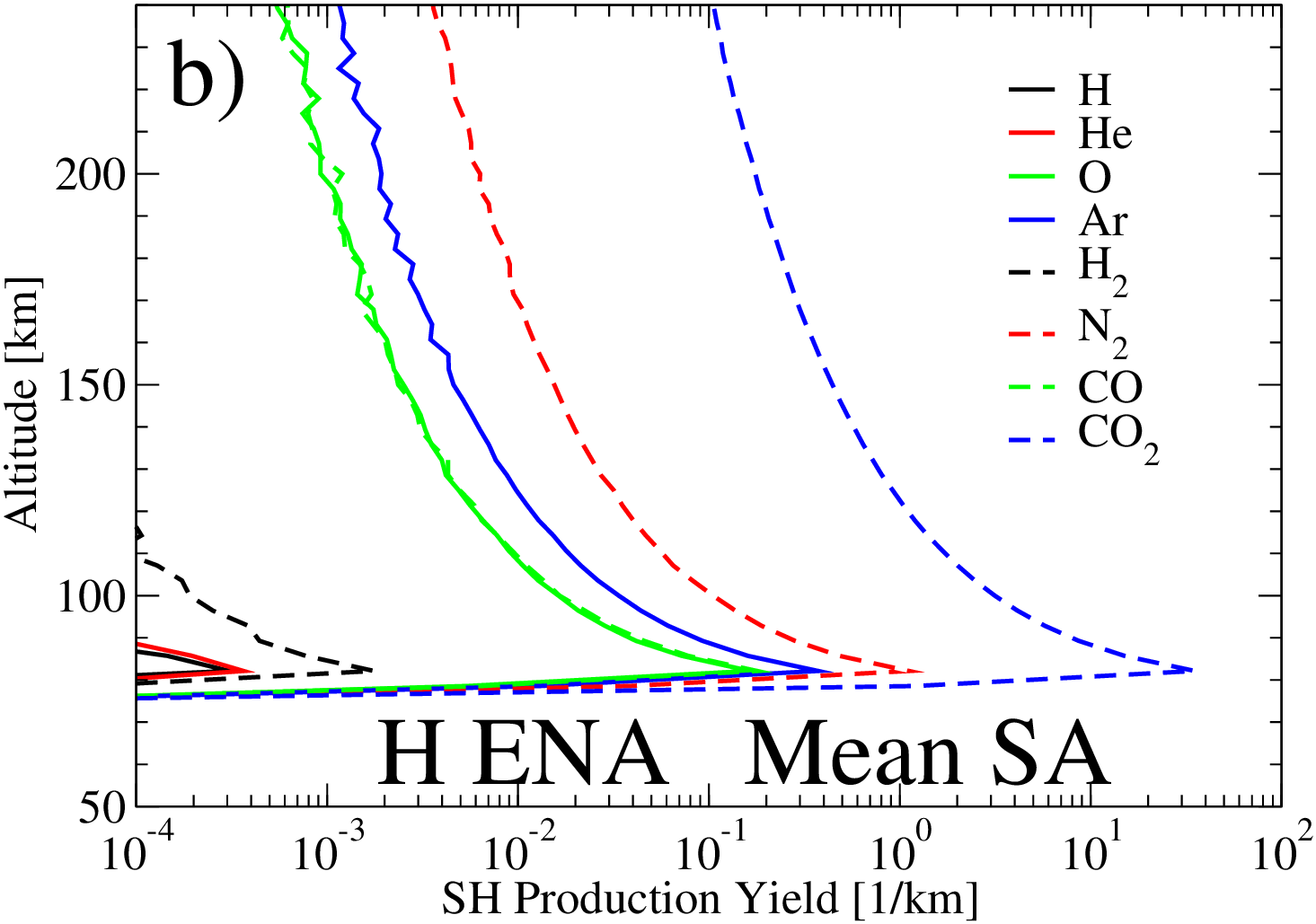}
\includegraphics[width=.35\linewidth]{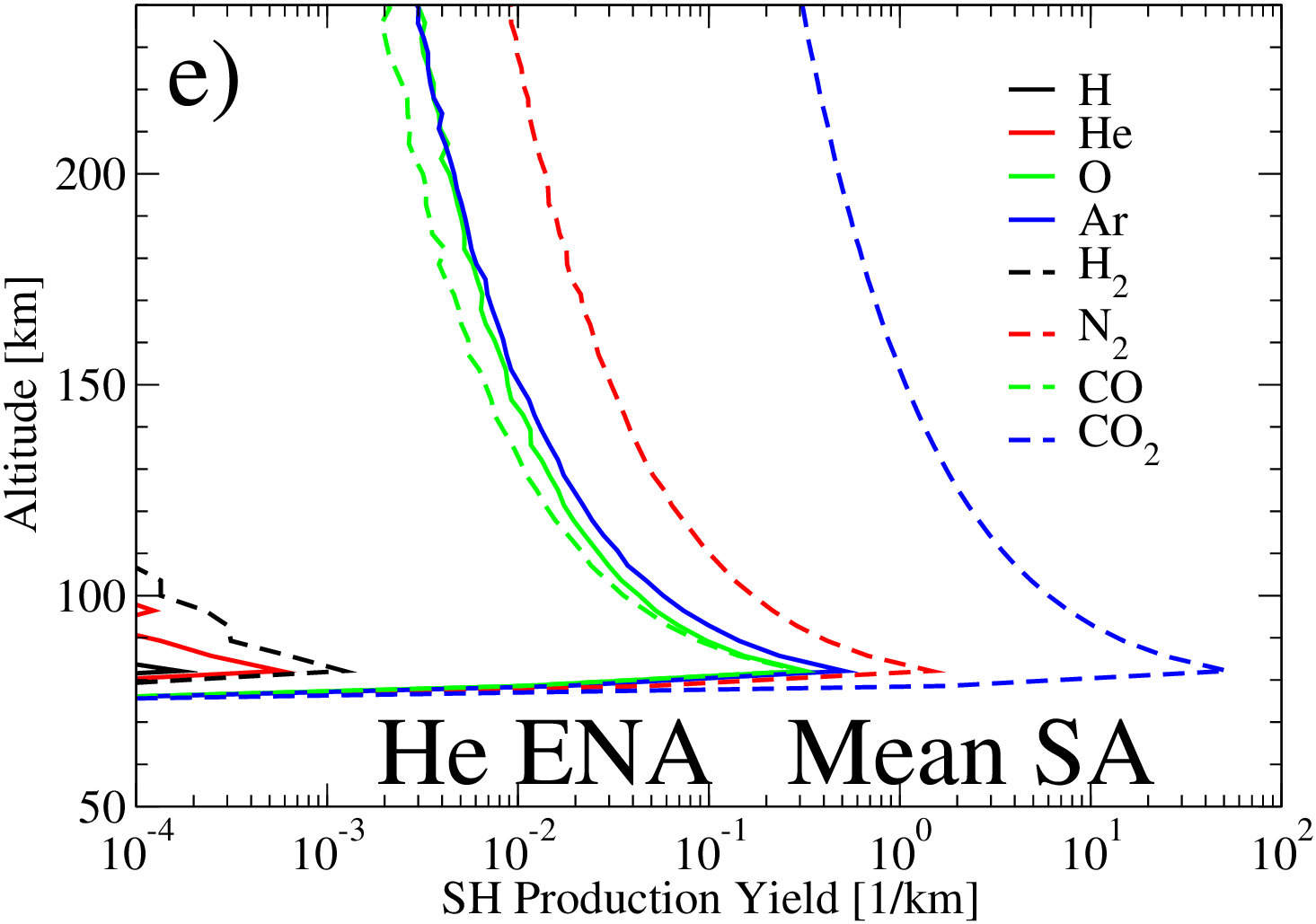} \\
	\vspace{25pt}
\includegraphics[width=.35\linewidth]{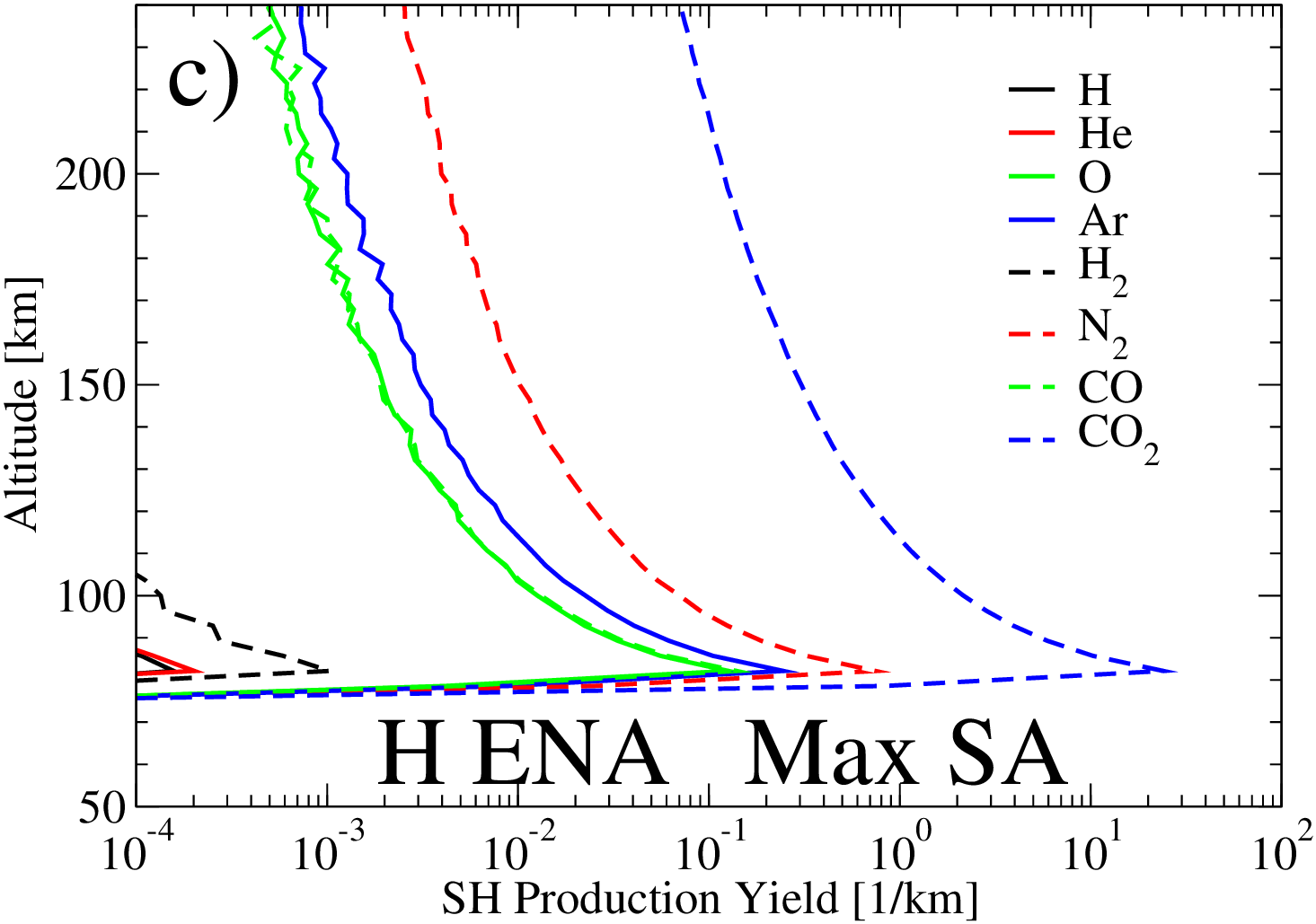}
\includegraphics[width=.35\linewidth]{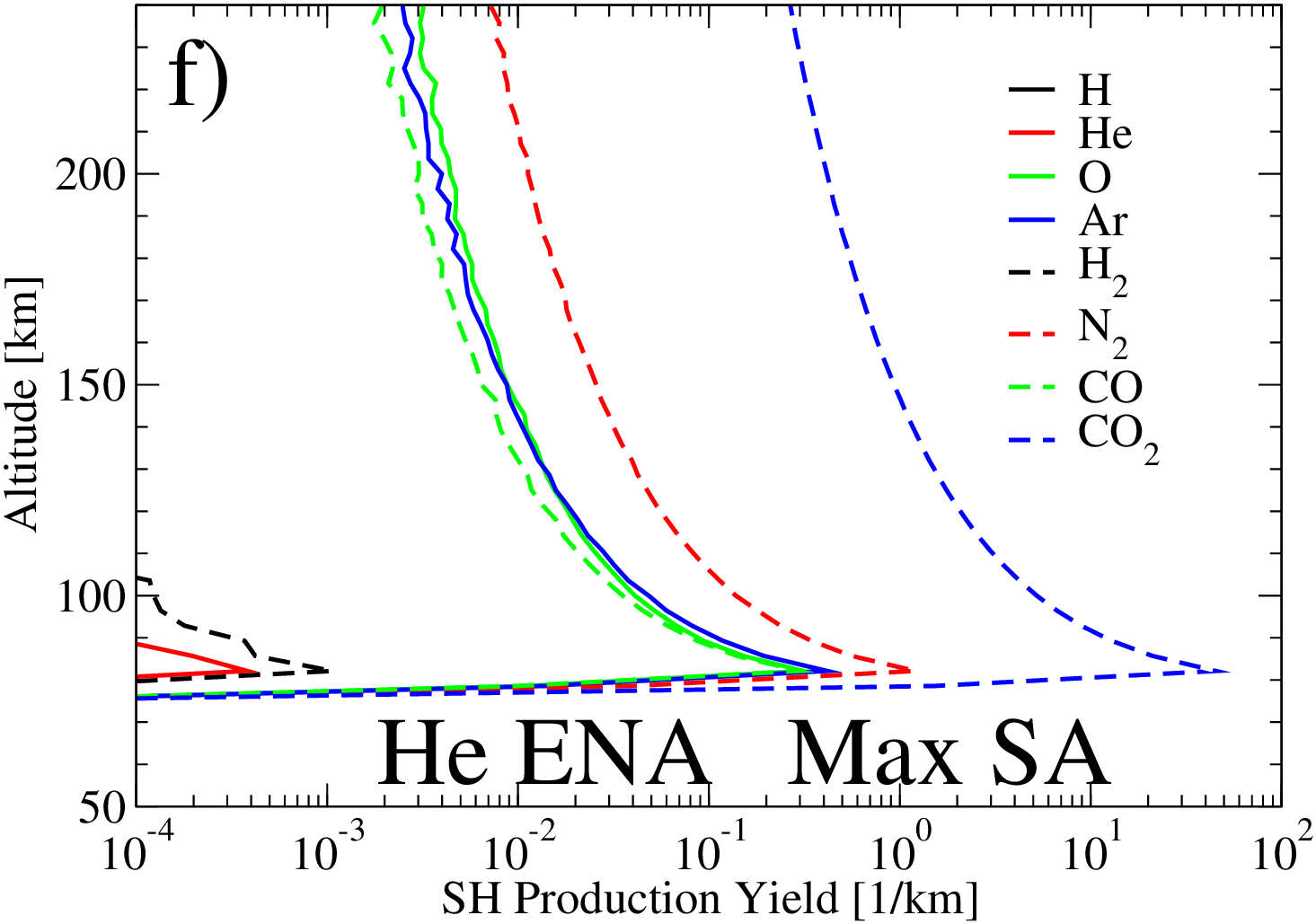} \\
\caption{\footnotesize SH atomic and molecular production yield due to precipitating hydrogen ENAs using the three atmosphere models for minimum a), mean b), and maximum SA c). The same data is shown for precipitating helium ENAs in figures d--f).}  
\label{SHA_Prod}
\end{figure*}

Production yields of SH atoms and molecules per precipitating particle were calculated as a function of altitude as
\begin{equation}
	Q(z) = \frac{N_{SH}(z)}{N \, \Delta z}
\end{equation}
where $N$ is the total number of incident ENAs, $N_{SH}(z)$ is the total number of nascent SH atoms/molecules created in the atmosphere layer at altitude $z$ with layer thickness $\Delta z$. 
Figure \ref{SHA_Prod} displays the SH atom/molecule production yields for major atmospheric species as a function of altitude. 
Additionally, to illustrate the dependence of production yields on solar conditions, Figure \ref{SHA_Prod} shows productions yields induced by both hydrogen and helium ENAs for all three neutral atmosphere models. 
For all atmospheric atoms and molecules, the profile shape for nascent SH atom/molecule production yields looks very similar with maximum production occurring deep in the atmosphere around 80 km.
{\color{black}
Even with the ensembles being initialized at an altitude of 700 km, maximum ENA production does not occur until the SW ions reach the altitude layer of 200 km as seen in Figure \ref{ENA_Prod}, and thermalization does not occur until the ENAs reach the altitude layers between 80--120 km, Figure \ref{Therm_Alt}.
The SH atom/molecule production rates, which are proportional to the local frequency of collisions between the precipitating energetic particles and atmospheric gas and molecules, seen in Figure \ref{SHA_Prod}, reflect the collision probabilities throughout all considered atmospheric layers with a steady increase with decreasing altitude and exponentially increasing gas densities. 
The maximum penetration depth of 80 km is reached where the ENAs transfer the remainder of their energy to the thermal gases, thus creating the production peaks seen in Figure \ref{SHA_Prod}.
Altitudes around 80 km can be reached only by the most energetic precipitating particles.
These high energy particles are involved the largest number of thermalizing collisions, the majority of which occur in the vicinity of the thermalization altitudes where the density of the atmospheric atoms and molecules is very high. 
The altitude layer of 80 km effectively stops all ENAs from penetrating any deeper into the atmosphere as the mean free path of ENAs in this layer is on the order of 1--10 cm.
}
For both hydrogen and helium ENAs incident on all atmospheric density models, the production yields for nascent SH H, H$_{2}$, and He are significantly lower than for other atmospheric species due to their small relative densities. 

\begin{figure}
	\vspace{20pt}
	\includegraphics[width=.85\linewidth]{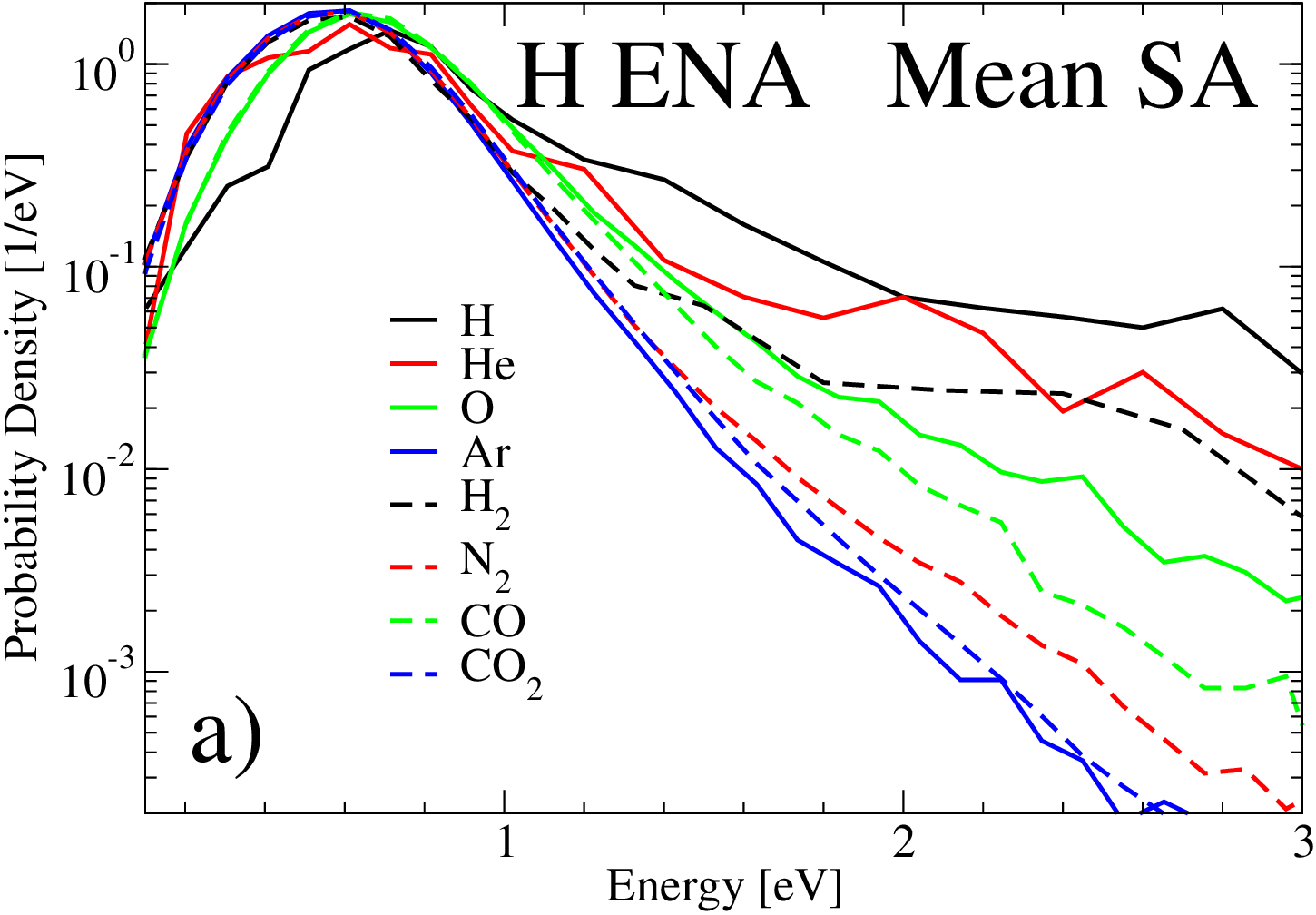}\hfill
	\vspace{30pt}
	\includegraphics[width=.85\linewidth]{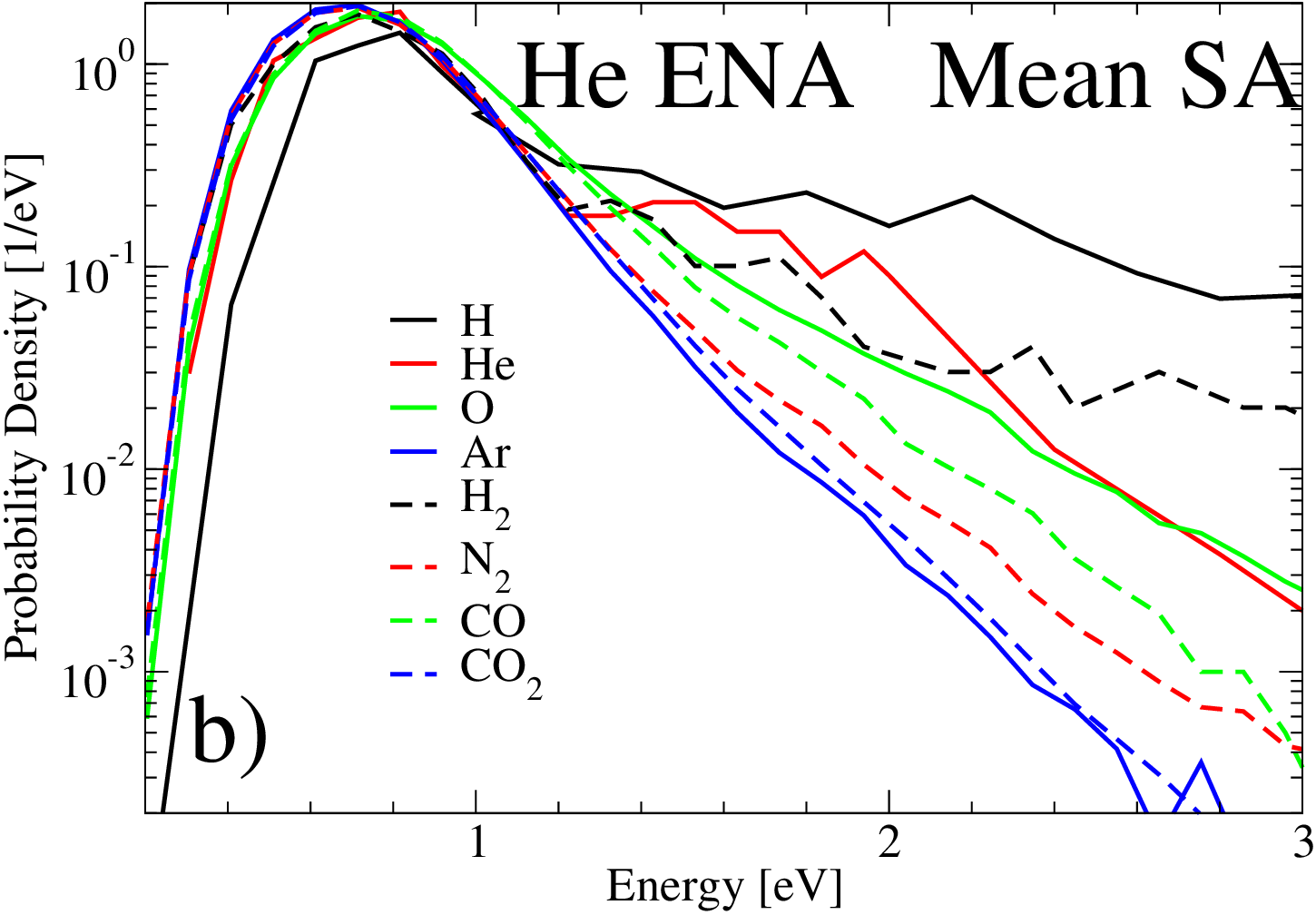} \\
	\caption{\footnotesize Nascent SH atomic and molecular energy distributions, normalized to unity, induced by collisions with hydrogen a) and helium b) ENAs using the mean SA atmosphere model.}
	\label{SHAM_EngDist}
\end{figure}

In addition to altitude dependent production yields, normalized energy distributions for nascent SH atoms and molecules were also calculated. 
The energy distributions for nascent SH atoms and molecules mostly reflect the energy-transfer processes of keV collisions. 
Because of this, the SH energy distributions were found to be near identical for the different atmosphere models and altitudes, and thus are presented in Figure \ref{SHAM_EngDist} as independent of altitude and for the mean SA atmosphere model only. 
The nascent energy distribution for SH atoms and molecules induced by helium ENAs is peaked at a slightly higher energy, 0.7 eV, as compared with the hydrogen ENAs, which is peaked at 0.5 eV. 
\begin{table}
\begin{center}
	\caption{Percentage of nascent secondary hot atoms and molecules capable of escaping\label{SH_can_Escape}}
	\begin{tabular}{ | c | c | c | c |}
\tableline
\tableline
     & \textbf{E}$_{esc}$ (eV) & \textbf{H ENA} (\%) & \textbf{He ENA} (\%) \\ 
\tableline
		H  			& 0.11 & 100   & 100 	\\ 	
		He  		& 0.44 & 80.18 & 93.75 \\ 	
		O  			& 1.8 & 1.94  & 3.34 	\\ 	
		Ar  		& 4.4 & 0     & 0 		\\ 	
		H$_{2}$ & 0.22 & 93.55 & 100 	\\ 	
		N$_{2}$ & 3.0 & 0.01  & 0.02 	\\ 	
		CO  		& 3.0 & 0.06  & 0.06 	\\ 	
		CO$_{2}$& 4.8 & 0     & 0 		\\ 	
\tableline
\end{tabular}
\end{center}
\tablecomments{SH atomic and molecular escape energies $E_{esc}$ calculated at 700 km.}
\end{table}
These differences in energy peaks are due in part by the larger average scattering angles of helium ENAs, Figure \ref{Rand_Angles}, as well as the mass ratio in Equation \ref{dE} favoring heavier projectiles for more energy transferred during a given collision.
To determine the fraction of the nascent SH atoms and molecules which have an energy above their respective escape energy, the distributions in Figure \ref{SHAM_EngDist} were integrated starting from the escape energy of the atom or molecule. 
Table \ref{SH_can_Escape} shows the escape energies of the SH atoms and molecules at 700 km as well as the percentage of nascent SH atoms and molecules, created by incident hydrogen and helium ENAs, which have energies above their respective escape energy. 
Due to their high masses and thus high escape energies, SH Ar and CO$_{2}$ do not have any realistic probability to escape as shown in Table \ref{SH_can_Escape}. 

Using the nascent SH atomic and molecular production yields along with the nascent energy distributions, SH atomic and molecular escape fluxes were estimated. 
Nascent SH atomic and molecular velocity directions were observed to be uniform across all simulations, so we assume initial velocity distributions are isotropic. 
These nascent isotropic distributions were used along with a simplified collision transparency formalism to estimate SH atomic and molecular escape fluxes induced by precipitating hydrogen and helium ENAs using the mean SA atmosphere model. 
In this simplified formalism, the ratio of escaping SH atomic or molecular fluxes, $\Phi_{esc}$, to incident ENA fluxes, $\Phi_{inc}$, may be written
\begin{equation}
\frac{\Phi_{esc}}{\Phi_{inc}} = \frac{(1+\alpha)}{2} \int\limits_{z_{min}}^{z_{max}} dz \, Q(z) \int\limits_{\epsilon_{esc}}^{\infty} d \epsilon \, \rho(\epsilon) \int\limits_{\sqrt{\displaystyle\frac{\epsilon_{esc}}{\epsilon}}}^{1} du \, T(u, z,\epsilon) 
\label{EscapeTrans}
\end{equation}
where SH atomic and molecular productions, $Q(z)$, are integrated over the altitude height, nascent energy distributions, $\rho(\epsilon)$, are integrated from the respective escape energy to infinity, and a conical component of the isotropic velocity distribution is integrated over a newly defined variable $u \equiv \cos \theta$, with $\theta$ being the standard polar angle, such that all velocity directions with upward components greater than escape velocity are considered.
The integral over this upward escape velocity cone also includes a collision transparency factor, $T(u,z,\epsilon)$, which gives the escape probability for a particle with velocity in the $u$ direction at altitude $z$ with energy $\epsilon$ and is defined as
\begin{equation}
T(u, z,\epsilon) = \exp \left( - \frac{1}{u}\int\limits_{z}^{z_{max}} \displaystyle\sum_{i} n_{i}(z') \sigma_{i}^{D}(\epsilon) \, dz' \right) 
\label{Trans}
\end{equation}
where $n_{i}(z)$ is the density of the $i^{th}$ atmospheric species at altitude $z$ and $\sigma_{i}^{D}(\epsilon)$ is the momentum-transfer cross section between the SH atom or molecule and the $i^{th}$ atmospheric species. 
The momentum-transfer (diffusion) cross section is defined using the differential cross section $|f(\epsilon,\theta)|^2$ as
\begin{equation}
	\sigma^{D}(\epsilon) = \int \left(1 - \cos \theta \right) |f(\epsilon,\theta)|^2 \, d \Omega 
\end{equation}
and is commonly used to describe processes involving energy transport \citep[]{Balakrishnan:1999,Zhang:2009,Bovino:2011}. 
The diffusion cross sections were used in Equation \ref{Trans} instead of total cross sections as they filter ultra small scattering angles which effectively transfer no energy to the target atoms \citep[]{Bovino:2011}. 
The factor $\alpha$ in Equation \ref{EscapeTrans} describes the fraction of nascent SH particles which have downward velocities and yet are reflected upward from the more dense gas layers due to large angle collisions. 
These upward reflections may contribute to the SH atomic and molecular escape fluxes. 
While these reflected SH particles may have different energy distributions than those shown in Figure \ref{SHAM_EngDist}, they are included since large reflection coefficients were observed for the incident keV/amu ENAs, discussed later in Section \ref{Section:Ref}. 
The reflection coefficients found in Section \ref{Section:Ref} range from 0.19 to 0.47 depending on the projectile ENA and the atmospheric condition.
For this work, the constant $\alpha = 0.2$ was used as it is the lower range of the reflection coefficients found in Section \ref{Section:Ref} and is a conservative estimate at the actual reflection value. 

The MC methods described is Section 4 were used to determine escape probabilities for SH helium created from precipitating hydrogen ENAs in a mean SA model atmosphere as a way to compare with the transparency formalism of Equation \ref{EscapeTrans}. 
Using 50,000 test particles with MC methods, the escape ratio was found to be 0.0170\% as compared to 0.0102\% calculated using Equation \ref{EscapeTrans} with $\alpha=0.2$. 
The transparency escape ratio differs by a percent error of 40\% from the {\color{black}MC} result making the simple transparency formalism described above a viable alternative to the time-expansive full {\color{black}MC} calculation. 
With the comparison made between {\color{black}MC} and collision transparency methods in good agreement, escape fluxes for all SH atomic and molecular species are shown with utilization of the collision transparency method. 

Realistic SH atomic and molecular escape fluxes were estimated using an average total incident ENA flux of $\Phi_{inc}^{tot}=9.41\times 10^{7}$ cm$^{-2}$ s$^{-1}$, calculated using an average SW speed of 437 km/s \citep[]{Reeves:2013} and an average SW ion density of 5 cm$^{-3}$ at 1 AU \citep[]{Phillips:1995}.
An average SW helium component of 2\% \citep[]{Aellig:2001} was used to determine the individual fluxes of precipitating hydrogen, $\Phi_{inc}^{H}=9.22\times 10^7$ cm$^{-2}$ s$^{-1}$, and helium, $\Phi_{inc}^{He}=1.88\times 10^6$ cm$^{-2}$ s$^{-1}$.
Table \ref{SH_Esc1} displays the escape probabilities calculated with Equation \ref{EscapeTrans} as well as estimated escape fluxes using the incident SW ion fluxes above.
{\color{black}Hydrogen} non-thermal escape fluxes are included in our table, although for the Mars atmosphere escape of {\color{black}hydrogen} atoms are governed by the Jeans thermal escape.  
In comparison, escape fluxes of neutral helium, due to collisions with hot oxygen created in dissociative recombination of O$_{2}^+$, were calculated to be 1.6$\times10^6$ cm$^{-2}$ s$^{-1}$ at minimum SA \citep[]{Bovino:2011}.
Oxygen escape fluxes, again due to {\color{black}dissociative recombination}, were calculated to range from $\times10^7$ cm$^{-2}$ s$^{-1}$ to $\times10^8$ cm$^{-2}$ s$^{-1}$ at minimum SA using several different collisional models \citep[]{Fox:2009,Fox:2010}.  
The escape fluxes estimated from the mechanism of SW ion precipitation are roughly three orders of magnitude smaller than the fluxes estimated due to hot oxygen generated in {\color{black}dissociative recombination} of O$_{2}^+$, yet still may be important in estimating long term atmospheric mass losses especially taking into account intensive fluxes of SW plasma during earlier periods in the Sun's history. 

\begin{table}
\begin{center}
	\caption{Secondary hot atomic and molecular escape probabilities and escape fluxes \label{SH_Esc1}}
\begin{tabular}{|c|c|c|c|c|}
\tableline
\tableline
\multicolumn{1}{|c|}{} &
\multicolumn{2}{c|}{\textbf{H ENA}} &
\multicolumn{2}{c|}{\textbf{He ENA}} \\
\cline{2-5}
\multicolumn{1}{|c|}{} &
\multicolumn{1}{c|}{\textbf{$\dfrac{\Phi_{esc}}{\Phi_{inc}}$}} & 
\multicolumn{1}{c|}{\textbf{$\Phi_{esc}$ $\left(\frac{1}{cm^{2} s}\right)$}} &
\multicolumn{1}{c|}{\textbf{$\dfrac{\Phi_{esc}}{\Phi_{inc}}$}} &
\multicolumn{1}{c|}{\textbf{$\Phi_{esc}$ $\left(\frac{1}{cm^{2} s}\right)$}} \\ [2.5ex]
\tableline
    H       &  4.54e-4 & 4.18e+4 & 9.42e-5 & 1.79e+2 \\
    He      &  1.02e-4 & 9.43e+3 & 4.83e-4 & 9.17e+2 \\
    O       &  5.57e-4 & 5.12e+4 & 3.62e-3 & 6.87e+3 \\
    Ar      &  0       & 0       & 0       & 0      \\
    H$_{2}$ &  8.13e-4 & 7.48e+4 & 1.52e-3 & 2.89e+3 \\
    N$_{2}$ &  1.04e-5 & 9.53e+2 & 3.62e-5 & 6.87e+1 \\    
    CO      &  7.40e-6 & 6.81e+2 & 2.23e-5 & 4.23e+1 \\
    CO$_{2}$&  0       & 0       & 0       & 0      \\
\tableline
    ALL     &  1.94e-3 & 1.79e+5 & 5.77e-3 & 1.10e+4 \\
\tableline
\end{tabular}
\end{center}
\tablecomments{SH escape probabilities were calculated using Equation \ref{EscapeTrans} while estimated escape fluxes were obtained using incident ENA fluxes of $\Phi_{inc}^{H}=9.22\times 10^7$ cm$^{-2}$ s$^{-1}$ and $\Phi_{inc}^{He}=1.88\times 10^6$ cm$^{-2}$ s$^{-1}$. Total escape probabilities and fluxes are shown in the last row for all SH atomic and molecular species.}
\end{table}


\subsection{ENA Reflection from the Mars Atmosphere and Escape}
\label{Section:Ref}

In addition to thermalizing and inducing SH atomic and molecular escape fluxes, the ENA test particles may also escape the atmosphere if they reach a height of 700 km with an upward velocity and an energy above their respective escape energy.  
Although all ensembles studied in this work are initialized directly incident to the surface of the planet with {\color{black}solar zenith angle} equal to 0$^{\circ}$, even for this case a significant percentage of all ensembles is reflected back upwards and escapes the planet.
\begin{table}
\begin{center}
	\caption{Reflection statistics for energetic neutral atoms\label{EnsembleTable}}
	\begin{tabular}{ | c | c | c | c | c | c | c | c |}
\tableline
\tableline
     & \textbf{CS} & \textbf{E$_{0}$} & \textbf{z$_{0}$} & \textbf{SA} & \textbf{P$_{R}$} & $\textbf{E}_{R}$ \textbf{(eV)} & $\textbf{N}_{R}$ \\ 
\tableline
		   & HS & PD    & PD     & Mean & 0.43 & 264 & 53 \\ 	
		   & QM & 1 keV & 200 km & Mean & 0.51 & 773 & 7549 \\ 	
\textbf{H}  & QM & PD    & PD     & Min  & 0.47 & 767 & 10989 \\ 	
		   & QM & PD    & PD     & Mean & 0.41 & 765 & 11091 \\ 	
		   & QM & PD    & PD     & Max  & 0.23 & 766 & 11003 \\ 	
\tableline
		   & HS & PD    & PD     & Mean & 0.14 & 245 & 66 \\	
		   & QM & 4 keV & 200 km & Mean & 0.23 & 2323 & 4757 \\ 	
\textbf{He} & QM & PD    & PD     & Min  & 0.24 & 2388 & 4906 \\ 
		   & QM & PD    & PD     & Mean & 0.22 & 2398 & 4835 \\	
		   & QM & PD    & PD     & Max  & 0.19 & 2384 & 4793 \\ 
\tableline
\end{tabular}
\end{center}
\tablecomments{Reflection statistics and ensemble parameters for collisional cross sections (CS) utilized, either hard sphere (HS) or quantum mechanical (QM), initial energy, E$_{0}$, initial altitude, z$_{0}$, and the solar activity (SA) Mars atmosphere model used during transport. PD is shown for parameters which utilized probability distributions for initial conditions. Ensemble averages for ENA reflection probability, P$_{R}$, reflected ENA energy, E$_{R}$, and number of collisions before being reflected from the atmosphere, N$_{R}$, are also shown in the table above.}
\end{table}

\begin{figure}
	\vspace{20pt}
	\includegraphics[width=.85\linewidth]{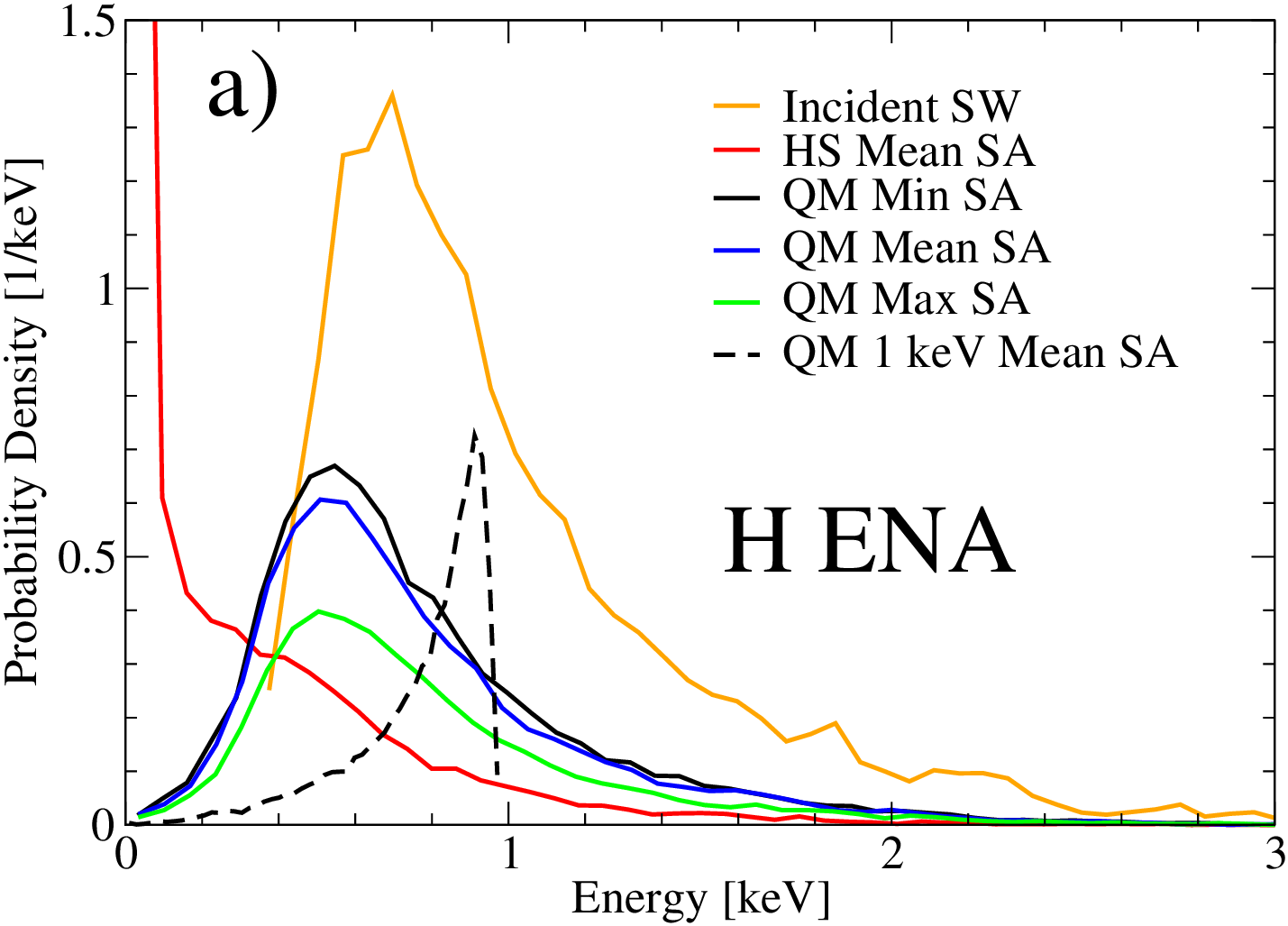}\hfill
	\vspace{30pt}
	\includegraphics[width=.85\linewidth]{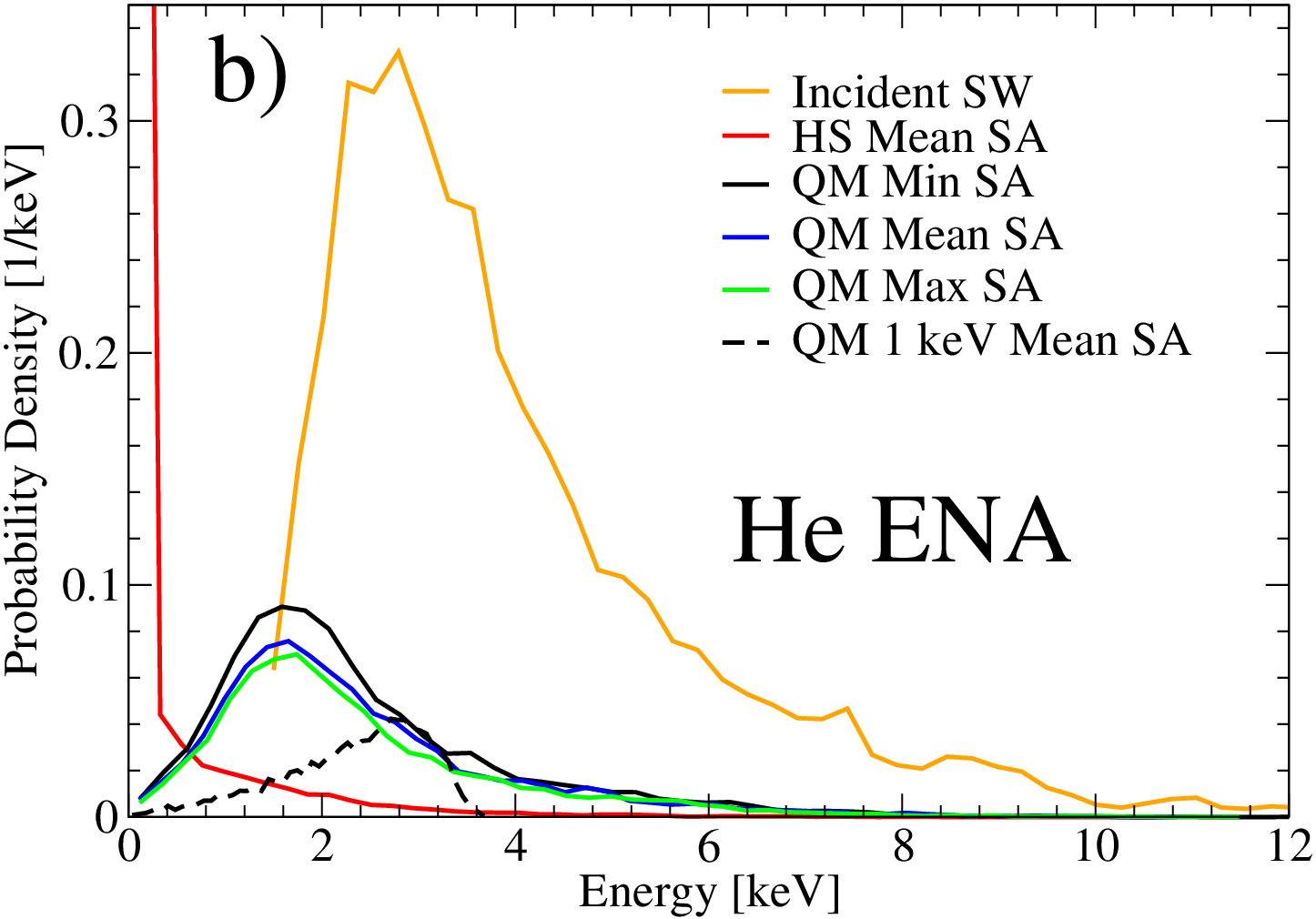} \\
	\caption{\footnotesize Energy distributions for hydrogen a) and helium b) ENAs reflected by the atmosphere of Mars. The incident energy distributions are shown, normalized to unity, as well as the reflected energy distribution for all ensembles studied in this work.}  
	\label{Escape_Dist}
\end{figure}

Table \ref{EnsembleTable} displays details of initial conditions for the different ensembles of precipitating particles used in our simulations as well as the reflection probability, average energy of reflected ENAs and average number of collisions before being reflected for each ensemble. 
In addition to Table \ref{EnsembleTable}, Figure \ref{Escape_Dist} displays the energy distributions for the reflected ENAs. 
For easy comparison, the initial incident energy distribution is shown, normalized to unity, for both hydrogen and helium ENAs in Figure \ref{Escape_Dist} while the reflected energy distributions for all ensembles are normalized to their reflection probability, $P_{R}$, in Table \ref{EnsembleTable}. 
Figure \ref{Escape_Dist} demonstrates how the reflected energy distributions are all shifted to lower energies due to the energy loss required from several large angle collisions to reflect the incident particles upward. 
The energy distributions from Figure \ref{Escape_Dist} also show how energetic the back reflected particles are, with the majority of escaping hydrogen having energies of 700 eV and reflected helium having energies of 2.3 keV. 
Our results for reflected hydrogen ENAs are similar to those previously reported by \cite{Kallio:2001} where it was observed that the reflection probability was 0.58. 
Additionally, \cite{Kallio:2001} reported that the ENAs which were reflected had an average energy of 470 eV which is more than half of their average initial energy of 800 eV. 
This signifies that if a particle does get reflected, it occurs quickly after becoming an ENA, before it gets very deep into the atmosphere where the transparency becomes low. 
This effect can be seen clearly from the mono-energetic ensembles in Figure \ref{Escape_Dist} which show peaks in the escaping energy probability around their initial energies of 1 and 4 keV. 
The HS ensembles are also informative as they display reflected energy distributions with significantly lower energies as compared to their QM counterparts. 
The atmosphere model used for the different ensembles also displays different escape probabilities, with the minimum SA having the highest escape probabilities and the maximum SA having the lowest. 
Differences in the energy distributions of reflected ENAs may be very useful for future missions to Mars as a mechanism for determining parameters of the upper atmosphere.  

\section{Conclusion}
Precipitation of ENAs, produced in the interaction between SW ions and atmospheric gas, has been investigated for the Mars atmosphere at different solar conditions.
For an accurate description of the energy relaxation process, the parameters for accurate descriptions of energy-momentum transfer in atomic and molecular collisions have been developed using both {\color{black}QM} methods and empirical models.
Properties of {\color{black}ENAs}, originating in the interaction between the SW ions and atmospheric gas, were calculated for the upper atmosphere of Mars using neutral atmosphere models for minimum, mean, and maximum solar activity.
{\color{black}MC} simulations were constructed to transport nascent {\color{black}ENAs} through the Martian atmosphere to determine properties of energy transfer, thermalization, production of {\color{black}SH} atoms and molecules, and reflection characteristics. 
Time-dependent energy distributions were obtained in addition to thermalization altitudes and atmospheric heating profiles.
Production rates and energy distributions for {\color{black}SH} atoms and molecules were extracted and utilized to determine induced atomic and molecular escape fluxes form Mars.
The information obtained from our {\color{black}MC} simulations demonstrates the need for accurate, energy-angular dependent cross sections in modeling the energy relaxation, sputtering and escape processes in planetary atmospheres as results obtained varied significantly between ensembles utilizing accurate cross sections and those which utilized isotropic, {\color{black}HS} cross sections. 

\begin{acknowledgments}
N. Lewkow and V. Kharchenko are grateful to NASA supporting our research via the grant NNX10AB88G.
\end{acknowledgments}

\bibliographystyle{apj}
\bibliography{Ref}

\begin{thebibliography}{43}
\expandafter\ifx\csname natexlab\endcsname\relax\def\natexlab#1{#1}\fi

\bibitem[{Aellig \& Lazarus(2001)}]{Aellig:2001}
Aellig, M., \& Lazarus, A. 2001, Geophysical Research Letters, 28, 2767

\bibitem[{Balakrishnan {et~al.}(1999)Balakrishnan, Kharchenko, \&
  Dalgarno}]{Balakrishnan:1999}
Balakrishnan, N., Kharchenko, V., \& Dalgarno, A. 1999, Journal of Chemical
  Physics, 103, 3999

\bibitem[{Barnett(1990)}]{Barnett:1990}
Barnett, C. 1990, in Atomic Data for Fusion, ed. H.~Hunter \& M.~Kirkpatrick,
  Vol.~1 (Oak Ridge National Laboratory: Controlled Fusion Atomic Data Center)

\bibitem[{Bondi(1964)}]{Bondi:1964}
Bondi, A. 1964, Physical Chemistry, 68, 441

\bibitem[{Bovino {et~al.}(2011)Bovino, Zhang, Gianturco, Dalgarno, \&
  Kharchenko}]{Bovino:2011}
Bovino, S., Zhang, P., Gianturco, F., Dalgarno, A., \& Kharchenko, V. 2011,
  Geophysical Research Letters, 38

\bibitem[{Cui {et~al.}(2012)Cui, Yelle, Strobel, M\"{u}ller-Wodarg, Snowden,
  Koskinen, \& Galand}]{Cui:2012}
Cui, J., Yelle, R., Strobel, D., {et~al.} 2012, Journal of Geophysical
  Research, 117

\bibitem[{Fox \& Ha\'{c}(2009)}]{Fox:2009}
Fox, J., \& Ha\'{c}, A. 2009, Icarus, 204, 527

\bibitem[{Fox \& Ha\'{c}(2010)}]{Fox:2010}
---. 2010, Icarus, 208, 176

\bibitem[{Fox \& Ha\'{c}(2014)}]{Fox:2014}
---. 2014, Icarus, 228, 375

\bibitem[{Fridman(2012)}]{Fridman:2012}
Fridman, A. 2012, Plasma Chemistry (Cambridge University Press)

\bibitem[{Gao {et~al.}(1990)Gao, Johnson, Hakes, Smith, \&
  Stebbings}]{Gao:1990}
Gao, R., Johnson, L., Hakes, C., Smith, K., \& Stebbings, R. 1990, Physical
  Review A, 41, 5929

\bibitem[{Gao {et~al.}(1988)Gao, Johnson, Schafer, Newman, Smith, \&
  Stebbings}]{Gao:1988}
Gao, R., Johnson, L., Schafer, D., {et~al.} 1988, Physical Review A, 38, 2789

\bibitem[{Gao {et~al.}(1989)Gao, Johnson, Smith, \& Stebbings}]{Gao:1989}
Gao, R., Johnson, L., Smith, K., \& Stebbings, R. 1989, Physical Review A, 40,
  4914

\bibitem[{Gealy \& Van~Zyl(1987)}]{Gealy:1987}
Gealy, M., \& Van~Zyl, B. 1987, Physical Review A, 36, 3091

\bibitem[{Greenwood {et~al.}(2000)Greenwood, Chutjian, \&
  Smith}]{Greenwood:2000}
Greenwood, J., Chutjian, A., \& Smith, S. 2000, The Astrophysical Journal, 529,
  605

\bibitem[{Hunten(1982)}]{Hunten:1982}
Hunten, D. 1982, Planetary Space Science, 30, 773

\bibitem[{Johnson {et~al.}(1989)Johnson, Gao, Hakes, Smith, \&
  Stebbings}]{Johnson:1989}
Johnson, L., Gao, R., Hakes, C., Smith, K., \& Stebbings, R. 1989, Physical
  Review A, 40, 4920

\bibitem[{Johnson {et~al.}(2008)Johnson, Combi, Fox, Ip, Leblanc, McGrath,
  Shematovich, Srobel, \& Waite~Jr.}]{Johnson:2008}
Johnson, R., Combi, M., Fox, J., {et~al.} 2008, Space Science Reviews, 139, 355

\bibitem[{Kallio {et~al.}(1997)Kallio, Luhmann, \& Barabash}]{Kallio:1997}
Kallio, E., Luhmann, J., \& Barabash, S. 1997, Journal of Geophysical Research,
  102, 183

\bibitem[{Kallio \& S.(2001)}]{Kallio:2001}
Kallio, E., \& S., B. 2001, Journal of Geophysical Research, 106, 165

\bibitem[{Kharchenko {et~al.}(1997)Kharchenko, Tharamel, \&
  Dalgarno}]{Kharchenko:1997}
Kharchenko, V., Tharamel, J., \& Dalgarno, A. 1997, Journal of Atmospheric and
  Solar-Terrestrial Physics, 59, 107

\bibitem[{Krasnopolsky(2002)}]{Krasnopolsky:2002}
Krasnopolsky, V. 2002, Journal of Geophysical Research, 107, 5118

\bibitem[{Krest'yanikova \& Shematovich(2005)}]{Krestyanikova:2005}
Krest'yanikova, M., \& Shematovich, V. 2005, Solar System Research, 39, 22

\bibitem[{Kusakabe {et~al.}(2002)Kusakabe, Buenker, \& Kimura}]{Kusakable:2002}
Kusakabe, T., Buenker, R., \& Kimura, M. 2002, in Atomic and Plasma-Material
  Interaction Data for Fusion, ed. R.~Clark, Vol.~10 (International Atomic
  Energy Agency)

\bibitem[{Ladd(2008)}]{Ladd:2008}
Ladd, J. 2008, PhD thesis, Colorado State University, Fort Collins, CO

\bibitem[{Lammer {et~al.}(2013)Lammer, Chassefi\'{e}re, Karatekin,
  Morschhauser, Niles, Mousis, Odert, M\:{o}stl, Breuer, Dehant, Grott,
  Gr\:{o}ller, Hauber, \& Pham}]{Lammer:2013}
Lammer, H., Chassefi\'{e}re, E., Karatekin, O., {et~al.} 2013, Space Science
  Review, 174, 113

\bibitem[{Lewkow {et~al.}(2012)Lewkow, Kharchenko, \& Zhang}]{Lewkow:2012}
Lewkow, N., Kharchenko, V., \& Zhang, P. 2012, The Astrophysical Journal, 756,
  57

\bibitem[{Lindsay \& Stebbings(2005)}]{Lindsay:2005}
Lindsay, B., \& Stebbings, R. 2005, Journal of Geophysical Research, 110

\bibitem[{Newman {et~al.}(1986)Newman, Chen, Smith, \& Stebbings}]{Newman:1986}
Newman, J., Chen, Y., Smith, K., \& Stebbings, R. 1986, Journal of Geophysical
  Research, 91, 8947

\bibitem[{Newman {et~al.}(1985)Newman, Smith, \& Stebbings}]{Newman:1985}
Newman, J., Smith, K., \& Stebbings, R. 1985, Journal of Geophysical Reseearch,
  90, 45

\bibitem[{Nitz {et~al.}(1987)Nitz, Gao, Johnson, Smith, \&
  Stebbings}]{Nitz:1987}
Nitz, D., Gao, R., Johnson, L., Smith, K., \& Stebbings, R. 1987, Physical
  Review A, 35, 4541

\bibitem[{Owen {et~al.}(1977)Owen, Biemann, Rushneck, Biller, Howarth, \&
  Lafleur}]{Owen:1977}
Owen, T., Biemann, K., Rushneck, D., {et~al.} 1977, Journal of Geophysical
  Research, 82, 4635

\bibitem[{Phillips {et~al.}(1995)Phillips, Bame, Barnes, Barraclough, Feldman,
  Goldstein, Gosling, Hoogeveen, McComas, Neugebauer, \& Suess}]{Phillips:1995}
Phillips, J., Bame, S., Barnes, A., {et~al.} 1995, Geophysical Research
  Letters, 22, 3301

\bibitem[{Reeves {et~al.}(2013)Reeves, Morley, \& Cunningham}]{Reeves:2013}
Reeves, G., Morley, S., \& Cunningham, G. 2013, Journal of Geophysical
  Research: Space Physics, 118, 1

\bibitem[{Sanchez-Lavega(2010)}]{Sanchez:2010}
Sanchez-Lavega, A. 2010, An Introduction to Planetary Atmospheres (Taylor \&
  Francis)

\bibitem[{Schafer {et~al.}(1987)Schafer, Newman, Smith, \&
  Stebbings}]{Schafer:1987}
Schafer, D., Newman, J., Smith, K., \& Stebbings, R. 1987, Journal of
  Geophysical Research, 92, 6107

\bibitem[{Schultz {et~al.}(2008)Schultz, Kristic, Lee, \&
  Raymond}]{Schultz:2008}
Schultz, D., Kristic, P., Lee, T., \& Raymond, J. 2008, The Astrophysical
  Journal, 678, 950

\bibitem[{Shematovich(2004)}]{Shematovich:2004}
Shematovich, V. 2004, Solar System Research, 38, 28

\bibitem[{Shematovich {et~al.}(2007)Shematovich, Tsvetkov, Krestyanikova, \&
  Marov}]{Shematovich:2007}
Shematovich, V., Tsvetkov, G., Krestyanikova, M., \& Marov, M. 2007, Solar
  System Research, 41, 103

\bibitem[{Smith {et~al.}(1996)Smith, Gao, Lindsay, Smith, \&
  Stebbings}]{Smith:1996}
Smith, G., Gao, R., Lindsay, B., Smith, K., \& Stebbings, R. 1996, Physical
  Review A, 53, 1581

\bibitem[{Steir \& Barnett(1956)}]{Stier:1956}
Steir, P., \& Barnett, C. 1956, Physical Review, 103, 896

\bibitem[{Wallace \& Hobbs(2006)}]{Wallace:2006}
Wallace, J., \& Hobbs, P. 2006, {Atmospheric Science, Second Edition: An
  Introductory Survey} (Academic Press)

\bibitem[{Zhang {et~al.}(2009)Zhang, Kharchenko, Jamieson, \&
  Dalgarno}]{Zhang:2009}
Zhang, P., Kharchenko, V., Jamieson, M., \& Dalgarno, A. 2009, Journal of
  Geophysical Research, 114

\end{thebibliography}

\end{document}